\documentclass[journal]{IEEEtran}
\usepackage{diagbox}
\usepackage[inline]{enumitem}
\usepackage{xcolor}
\usepackage{amsmath,amsfonts}
\usepackage{algorithm}
\usepackage[group-separator={,},group-minimum-digits=4]{siunitx}
\usepackage{threeparttable}
\usepackage{array}
\usepackage[caption=false,font=normalsize,labelfont=sf,textfont=sf]{subfig}
\usepackage{textcomp}
\usepackage{stfloats}
\usepackage{url}
\usepackage{verbatim}
\usepackage{graphicx}
\usepackage{algpseudocode}
\usepackage{cite}
\usepackage{balance}
\usepackage{mathrsfs}
\usepackage[colorlinks=true,
            linkcolor=blue,
            anchorcolor=blue,
            citecolor=blue]{hyperref}

\begin{document}

\title{
 {Disco Intelligent Omni-Surfaces: 360$^\circ$ Fully-Passive Jamming Attacks}
 
} 
\author{ 
{
    Huan~Huang,~\textit{Member,~IEEE}, 
    Hongliang~Zhang,~\textit{Member,~IEEE},
    Jide~Yuan,~\textit{Member,~IEEE},
    Luyao~Sun,
    Yitian~Wang, 
    Weidong~Mei,~\textit{Member,~IEEE}, 
    Boya~Di,~\textit{Member,~IEEE},
    Yi~Cai,~\textit{Senior~Member,~IEEE}, 
    and~Zhu~Han~\textit{Fellow,~IEEE}
}
\thanks{  

A portion of this work was published in~\cite{MyDIOSICC24}.

H.~Huang, J.~Yuan, L.~Sun, Y.~Wang, and Y.~Cai are with the School of Electronic and Information Engineering, Soochow University, Suzhou, Jiangsu 215006, China 
(e-mail: hhuang1799@gmail.com, jide\_yuan@suda.edu.cn, lysun02@163.com, ytwang41@stu.suda.edu.cn, yicai@ieee.org). 

H.~Zhang and B.~Di are with the 
School of Electronics, Peking University, Beijing 100871, China 
(email: hongliang.zhang92@gmail.com, diboya@pku.edu.cn). 

W.~Mei is with the National Key Laboratory of Wireless Communications,
University of Electronic Science and Technology of China, Chengdu 611731, China (e-mail: wmei@uestc.edu.cn).

Z.~Han is with the Department of Electrical and Computer Engineering,
University of Houston, Houston, TX 77004 USA,
and also with the Department of Computer Science and Engineering, 
Kyung Hee University, Seoul, South Korea, 446-701.  
(email: hanzhu22@gmail.com).
}
}
\maketitle

\begin{abstract}
Intelligent omni-surfaces (IOSs) with 360$^{\rm o}$ electromagnetic radiation significantly improves the performance of wireless systems, 
while an adversarial IOS also poses a significant potential risk for physical layer security.
In this paper, we propose a ``DISCO'' IOS (DIOS) based fully-passive jammer (FPJ) that can launch omnidirectional fully-passive jamming attacks.
In the proposed DIOS-based FPJ, the interrelated refractive and reflective (R\&R) coefficients of the adversarial IOS are randomly generated, acting like a ``DISCO ball'' that distributes wireless energy radiated by the base station.
By introducing active channel aging (ACA) during channel coherence time, the DIOS-based FPJ can perform omnidirectional fully-passive jamming without neither jamming power nor channel knowledge of legitimate users (LUs).
To characterize the impact of the DIOS-based PFJ, we derive the statistical characteristics of DIOS-jammed channels based on two widely-used IOS models, i.e., the constant-amplitude model and the variable-amplitude model.
Consequently, the asymptotic analysis of the ergodic achievable sum rates under the DIOS-based omnidirectional fully-passive jamming is given based on the derived stochastic characteristics for both the two IOS models.
Based on the derived analysis, the omnidirectional jamming impact of the proposed DIOS-based FPJ implemented by a constant-amplitude IOS does not depend on either the quantization number or the stochastic distribution of the DIOS coefficients, 
while the conclusion does not hold on when a variable-amplitude IOS is used.
Numerical results based on one-bit quantization of the IOS phase shifts are provided to 
verify the effectiveness of the derived theoretical analysis. 
The proposed DIOS-based FPJ can not only launch omnidirectional fully-passive jamming,
but also improve the jamming impact by about 55\% at 10 dBm transmit power per LU.
\end{abstract}


\section{Introduction}\label{Intro}
Due to the inherent broadcast and superposition properties of wireless channels, 
wireless systems are vulnerable to malicious physical-layer attacks such as physical-layer jamming,
which is a type of denial-of-service (DoS) attacks~\cite{PLSsur1,DoSsur1,AntiJammingSurv}.
In wireless systems, physical-layer jamming can be easily launched by an active jammer (AJ),
which inflicts intentional jamming/interference attacks to block the wireless communication 
between the access point (AP) and the legitimate users (LUs).
Typical AJs can be divided into the following categories:
1) constant AJs, 2) intermittent AJs~\cite{intermittentAJ}, 3) reactive AJs~\cite{reactiveAJ}, and 4) adaptive AJs~\cite{adaptiveAJ}.
These AJs have the inherent energy limitation because 
the AJs require to broadcast intentional jamming/interference,
such as pseudorandom noise or modulated Gaussian waveforms, over an open wireless channel.
Therefore, an important metric of AJs is  the development of strategies to maximize the duration and area of effective jamming 
while minimizing jamming power.
However, the AJs inevitably consume a certain amount of jamming energy 
to prevent LUs from communicating with legitimate APs.

Recently, reconfigurable intelligent surfaces (RISs) have attracted increasing attention as a promising candidate technology for the future sixth generation (6G) wireless communications~\cite{IOSImple,RISSur, RISSur1}.
Existing works mainly focus on the use of RISs to improve system performance, 
e.g. minimizing energy efficiency~\cite{MyMultiRIS,EE1} or maximizing spectrum efficiency~\cite{SE1,DIOSNearFarModel},
where the RIS coefficients should be carefully designed according to the channel state information (CSI).
Unlike legitimate RISs, adversarial RISs, the illegitimate utilization of RISs~\cite{illIRS} 
poses a significant detrimental impact on wireless systems, which needs to be given increasing attention. 

\subsection{Related Works}
Some existing works have focused on the detrimental impact of adversarial RISs.
For example, the work~\cite{PassJamSU} has reported an adversarial RIS-based passive jammer (PJ) 
that destructively adds the reflected path signal to the direct path signal to minimize the received power, i.e., to minimize the signal-to-noise ratio (SNR). 
Then the communications between the legitimate AP and its LU in the single-user multiple-input single-output (SU-MISO) system are blocked.
Moreover, the authors in~\cite{PAdRIS} investigated the use of an adversarial RIS to jam an multi-user multiple-input single-output (MU-MISO) system.
Similarly, the attacker must carefully calculate the reflective coefficients of the adversarial RIS in order to minimize the sum rate,
i.e., to minimize the signal-to-interference-plus-noise ratio (SINR).
Although these adversarial RIS can jam LUs without consuming jamming power, 
CSI of all wireless channels, including the wireless channels between legitimate APs and LUs, must be known at the adversarial RIS. 
Due to the passive nature of RISs, it is unrealistic to assume that the illegitimate RIS knows the CSI, 
especially the CSI of the wireless channels between legitimate APs and LUs. 

To address the limitation of adversarial RISs in acquiring the CSI,
fully-passive jammers (FPJs) have been proposed~\cite{MyDRIS1,MyDRIS2,MyDRIS5}, 
which can launch jamming attacks without relying on either jamming power or CSI.
The concept of FPJ was first proposed in~\cite{MyDRIS1}, 
where an adversarial RIS with random and time-varying reflective coefficients acts like a ``DISCO ball'' and is therefore called a DISCO RIS (DRIS)~\cite{MyDRIS2}.
The use of DRIS causes active channel aging (ACA) and then fully-passive jamming is generated~\cite{MyDRIS5}.
Note that the ACA is different from traditional channel aging (CA)~\cite{ChanAge} 
that ioccurs due to time variations in RF propagation 
and computational delays between the moment the wireless channels are acquired at the AP 
and when they are applied for precoding.
Moreover, some works investigated the introduction of DIRSs to break key consistency in channel reciprocity-based key generation~\cite{OtherDIRS1,OtherDIRS2,OtherDIRS3}
or to break channel reciprocity-based communications~\cite{OtherDIRS4} in time division duplex (TDD) wireless systems.
For more clarity, we list and compare these adversarial RIS-based jamming schemes in Table~\ref{tab1add}.
\begin{table*}
    \footnotesize
    \centering
    \caption{Comparison of Different Jammers}
    \label{tab1add}
    \begin{threeparttable}
        \begin{tabular}{| m{2.5cm}<{\centering} | m{3.5cm}<{\centering} | m{2.2cm}<{\centering} | m{2.4cm}<{\centering} | m{2.2cm}<{\centering} |}
    \hline
    Reference                                                                    &Mechanism                                                &Jamming power    &Channel knowledge   &Jamming area\\
    \hline
    \cite{PassJamSU,PAdRIS}                                                      &Optimize RIS coefficients to minimize SNR or SINR           &Not Required     &Required            &Reflective side\\
    \hline
    \cite{MyDRIS1,MyDRIS2,MyDRIS5,OtherDIRS1,OtherDIRS2,OtherDIRS3,OtherDIRS4}   &Break channel reciprocity in TDD systems                 &Not Required     &Not Required        &Reflective side\\                                 
    \hline
    \end{tabular}
    \end{threeparttable}
\end{table*}

It can be seen from Table~\ref{tab1add}, although the DRIS-based FPJs can launch fully-passive jamming attacks without relying on either jamming power or LU channel knowledge, 
they can only jam the LUs located on the reflective side of the DRIS.
Namely, there are blind jamming areas, 
where the LUs located on the refractive side of the DRIS are completely unable to be jammed by the DRIS-based FPJ.
Immediately following the studies on RISs, 
intelligent omni-surfaces (IOSs) are being introduced into wireless communications 
to achieve 360$^\circ$ performance improvement by enabling the simultaneous reflection and refraction~\cite{IOS1,STAR1,STAR2,STAR3,STAR4,IOS3,IOS2} .
It should be noted that an IOS is not the same as two independent reflective RISs back to back~\cite{IOS1,IOS2,IOS3},
because there is an additional constraint between the refractive and reflective (R\&R)
coefficients of each IOS element.
Due to this additional constraint, an IOS can not be directly introduced into the DRIS-based FPJ~\cite{MyDRIS1,MyDRIS2,MyDRIS5} to implement omnidirectional fully-passive jamming.
Considering this constraint of an IOS, the work~\cite{MyDIOSICC24} first proposed the concept of omnidirectional FPJ, which introduces a DISCO IOS (DIOS) to 
implement 360$^\circ$ fully-passive jamming attacks.
However, the authors in~\cite{MyDIOSICC24} only
demonstrated the impact of the DIOS-based FPJ on an MU-MISO system through simulations, 
without providing a theoretical analysis.
\subsection{Contributions and Organization}
In this work, we propose a DIOS-based FPJ that can launch 360$^\circ$ fully-passive jamming attacks 
without relying on either jamming power or LUs' channel knowledge.
To quantify the impact of these omnidirectional fully-passive jamming attacks, the quantitative analysis is performed. 
The main contributions are summarized as follows:
\begin{itemize}
\item  We investigate the downlink rate of an MU-MISO system jammed by the proposed DIOS-based FPJ.
In the proposed DIOS-based FPJ, the DIOS remains ``silent'' during each pilot transmission (PT) phase,
where the term ``silent'' refers to the wireless signals being perfectly absorbed by
the adversarial DIOS~\cite{RISSilent}. Then, the DIOS randomly changes its R\&R
coefficients during the subsequent data transmission (DT) phase. 
In other words, the DIOS with  random and time-varying R\&R coefficients acts like a ``DISCO ball'' that distributes the AP transmit power in random directions.
As a result, the AP-LU channels change rapidly, causing serious inter-user interference, referred to as  active channel age (ACA).
\item 
Two widely-used IOS models, i.e., the constant-amplitude IOS model and the variable-amplitude IOS model, 
are introduced into the investigation of the DIOS-based FPJ.
In the two IOS models, the R\&R phase shifts of the IOS elements are discrete and interrelated. 
In the constant-amplitude IOS model, we assume that the R\&R amplitudes of each IOS element are constant and equal.
Yet, in the variable-amplitude IOS model, the R\&R amplitudes of each IOS element are assumed to be dependent and different for different R\&R phase shifts,
and the R\&R amplitudes of each IOS element 
are also not equal, alternating due to the energy conservation constraint.
For both constant-amplitude and variable-amplitude IOS models, we perform the proposed DIOS-based FPJ under the constraint that the R\&R coefficients are related.
\item To quantify the impact of the omnidirectional fully-passive jamming, 
we give the asymptotic analysis of the achievable
sum rates under the above two IOS assumptions, i.e., the constant-amplitude DIOS assumption and the variable-amplitude DIOS assumption.
First, the statistical characteristics of the DIOS-jammed channels are given for both the two DIOS models.
Then, the lower bounds of the downlink rates are derived for 
both the refractive-side LUs and the reflective-side LUs based on the derived statistical characteristics.
\item Based on the detailed asymptotic analysis, we present some unique properties of the proposed DIOS-based FPJ. 
For instance, the jamming impact is not dependent on either the quantization bits or the distribution of the R\&R phase shifts when the constant-amplitude DIOS is exploited. 
However, when the variable-amplitude DIOS is used, the jamming impact depends on the quantization bits and the distribution. 
Since the jamming impacts on the refractive-side LUs and the reflective-side LUs 
are related by energy conservation,
 we can carefully design a distribution
to balance the impacts of the DIOS-based omnidirectional fully-passive jamming attacks on the refractive-side LUs and the reflective-side LUs.
\end{itemize}

The rest of this paper is organized as follows.  
In Section~\ref{Princ}, the downlink of an MU-MISO system jammed by the proposed DIOS-based FPJ is first modeled, where the performance metric used to quantify the omnidirectional jamming impact is given.
Then, all wireless channels involved are modeled.
In Section~\ref{DIOSFPJImpact}, the statistical characteristics of the time-varying R\&R DIOS-jammed channels are derived based on two widely-used IOS models, i.e., the constant-amplitude model and the variable-amplitude model. 
Then, the asymptotic analysis of the proposed DIOS-based FPJ is performed,
where the lower bounds of ergodic achievable R\&R sum rates are derived.
Simulation results 
 are presented in Section~\ref{ResDis} 
 to demonstrate the effectiveness of the derived asymptotic analysis and the jamming impact of the proposed DIOS-based FPJ.  
Finally, the main conclusions are summarized in Section~\ref{Conclu}.

\emph{Notation:} We employ bold capital letters for a matrix, e.g., ${\bf{W}}_{\rm{ZF}}$, lowercase bold letters for a vector, e.g., $\boldsymbol{w}_{{\rm {ZF}},k}$, and italic letters for a scalar, e.g., $K$. The superscripts $(\cdot)^{T}$ and $(\cdot)^{H}$ represent the transpose and the Hermitian transpose, respectively, and the symbols $\|\cdot\|$ and $|\cdot|$  represent the Frobenius norm and the absolute value, respectively. 

\section{System Description}\label{Princ}
In this section, we first describe an MU-MISO system under the jamming attacks launched by the DIOS-based FPJ.
Then, all wireless channels involved are built.

\subsection{Disco IOS Based Fully-Passive jammer}\label{DIOSFPJ}
\begin{figure}[!t]
    \centering
    \includegraphics[scale=0.7]{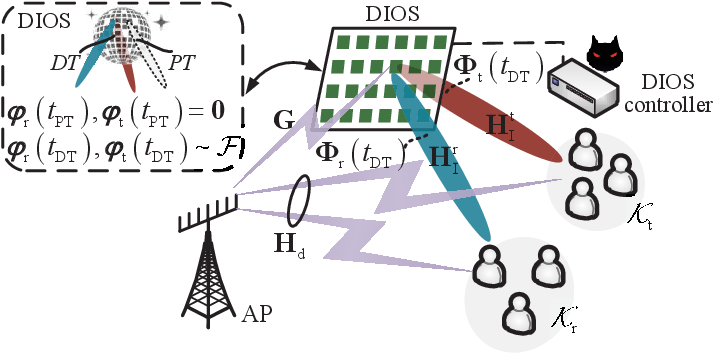}
    \caption{An illustration of an MU-MISO system jammed by DISCO intelligent omni-surface (DIOS) based fully-passive jamming attacks,
    where the DIOS is turned off during the \emph{pilot transmission (PT)} and then turned on during\emph{data transmission (DT)} phases.}
    \label{fig1}
\end{figure}

Fig.~\ref{fig1} schematically shows an MU-MISO system attacked by the proposed DIOS-based FPJ,
where the DIOS-based FPJ launches omnidirectional fully-passive jamming attacks without relying on jamming power and CSI.
We assume that the legitimate AP equipped with an $N_{\rm A}$-element uniform linear array (ULA) communicates with total $K$ LUs denoted by ${\cal K} = \left\{1,\cdots, K\right\}$.
Furthermore, 
we assume that  $K_{{\rm t}}$ LUs termed as ${\cal K}_{\rm t} = \left\{1,\cdots, K_{\rm t}\right\}$ 
and $K_{{\rm r}}$ LUs termed as ${\cal K}_{\rm r} = \left\{1,\cdots, K_{\rm r}\right\}$ are respectively located on the refractive and reflective (R\&R) side of the DIOS, 
where $K = K_{{\rm r}} + K_{{\rm t}}$.
Similar to the deployment in~\cite{MyDIOSICC24,MyDRIS1,MyDRIS2,MyDRIS5}, the DIOS is implemented close to the AP.

Generally, during the channel coherence time in an MU-MISO system, the AP designs the transmit beamforming used in the \emph{DT} phase
based on the CSI acquiring from the \emph{PT} phase.
Furthermore, we assume that the length of a \emph{DT} phase is $C$ times longer than that of a \emph{PT} phase,
i.e., $T_{\rm D} = CT_{\rm P}$.
Similar to the setting in~\cite{MyDRIS2,MyDRIS5}, 
the DIOS is turned off during the \emph{PT} phase and then turned on during the \emph{DT} phase with random and time-varying R\&R coefficients,
where the period during which the R\&R coefficients are changing is about the same as the length of the \emph{PT} phase $T_{\rm P}$.
Mathematically, we denote R\&R passive beamforming as ${\bf \Phi}_{{{\rm t}}}(t) = {\rm{diag}}\left({\boldsymbol{\varphi}}_{{{\rm t}}}(t)\right)$
and ${\bf \Phi}_{\rm r}(t) = {\rm{diag}}\left({\boldsymbol{\varphi}}_{\rm r}(t)\right)$~\cite{TJCuiIOS,IOS1,STAR1}, 
where the the random and time-varying R\&R vectors are respectively expressed as
\begin{equation} 
    {\boldsymbol{\varphi}}_{\rm t}(t) = \left[ {{\alpha^{\rm{t}}_{1}}\!\left( t \right){e^{j{\varphi^{\rm{t}}_{1}}\left( t \right)}}, \cdots ,{\alpha^{\rm{t}}_{{N_{\!\rm{D}}}}}\!\!\left( t \right){e^{j{\varphi^{\rm{t}}_{{N_{\!\rm{D}}}}}\left( t \right)}}} \right],
    \label{PBFT}
\end{equation}
and
\begin{equation} 
    {\boldsymbol{\varphi}}_{\rm r}(t) = \left[ {{\alpha^{\rm{r}}_{1}}\!\left( t \right){e^{j{\varphi^{\rm{r}}_{1}}\left( t \right)}}, \cdots ,{\alpha^{\rm{r}}_{{N_{\!\rm{D}}}}}\!\!\left( t \right){e^{j{\varphi^{\rm{r}}_{{N_{\!\rm{D}}}}}\left( t \right)}}} \right].
    \label{PBFR}
\end{equation}
In~\eqref{PBFT} and~\eqref{PBFR}, the R\&R amplitudes ${\alpha^{\rm{t}}_{m}}(t)$ and ${\alpha^{\rm{r}}_{m}}(t)$ of the $m$-th DIOS element 
($m = 1, \cdots, N_{\rm D}$) are a function of their corresponding R\&R phase shifts ${\varphi^{\rm{t}}_{m}}(t)$ and ${\varphi^{\rm{r}}_{m}}(t)$~\cite{IOS1,IOS3}.
Furthermore, ${\alpha^{\rm{t}}_{m}}(t)$ and ${\alpha^{\rm{r}}_{m}}(t)$ satisfy the energy conservation constraint, i.e.,  $\left|{\alpha^{\rm{t}}_{m}}(t)\right|^2 + \left|{\alpha^{\rm{r}}_{m}}(t)\right|^2 = 1$~\cite{STAR1}.

In practice, an IOS is an ultra-thin surface 
composed of multiple sub-wavelength elements whose  
 R\&R coefficients are controlled by simple programmable PIN or varactor diodes~\cite{IOSImple}.
We assume that the programmable PINs are used to implement the DIOS, whose ON/OFF behavior only allows for the creation of discrete phase shifts.
Therefore, we denote the $b$-bit refractive phase set as ${\Theta _{\rm{t}}} = \left\{ {\theta _1^{\rm{t}}, \cdots \theta _{{2^{{b}}}}^{\rm{t}}} \right\}$
and the $b$-bit reflective phase set as ${\Theta _{\rm{r}}} = \left\{ {\theta _1^{\rm{r}}, \cdots \theta _{{2^{{b}}}}^{\rm{r}}} \right\}$, respectively.
Note that an IOS has an additional constraint compared to a RIS
that the R\&R phase shifts ${\varphi^{\rm{t}}_{m}}(t)$ and ${\varphi^{\rm{r}}_{m}}(t)$ are inter-related~\cite{IOS2,IOS3}.
Namely, ${\varphi^{\rm{t}}_{m}}(t)$ is $\theta^{\rm t}_{m} \in {\Theta _{\rm{t}}}$ if ${\varphi^{\rm{r}}_{m}}(t)$ takes $\theta^{\rm r}_{m} \in {\Theta _{\rm{r}}}$. 
Moreover, the R\&R amplitudes are assume to take from the sets ${\Lambda}_{\rm t} = \left\{\xi^{\rm t}_1,\cdots,\xi^{\rm t}_{2^b}\right\}$
and ${\Lambda}_{\rm r} = \left\{\xi^{\rm r}_1,\cdots,\xi^{\rm r}_{2^b}\right\}$, 
where $\left|\xi^{\rm r}_s\right|^2 + \left|\xi^{\rm t}_s\right|^2 = 1, s= 1,\cdots, 2^b$.

\underline{\emph{Transmit Beamforming Design:}}
In the $PT$ phase of each channel coherence time, 
the CSI estimated by using methods such as the least squares (LS) algorithm~\cite{DaipartI} is expressed as
${\bf H}^{H}_{\rm PT} = {\bf H}^{H}_{\rm d} = \left[{{\boldsymbol{h}}_{{\rm d},1}},\cdots,{{\boldsymbol{h}}_{{\rm d},K}}\right]^{H} \!\in\! {\mathbb{C}}^{K \times {N_{\! \rm A}}}$,
i.e., the overall direct channel between the $K$ LUs and the AP.
Based on the CSI of ${\bf H}_{\rm{PT}}$, the AP then designs the transmit beamforming used to send signals to the LUs during the following \emph{DT} phase.
Without loss of generality, we assume that the AP uses zero-forcing (ZF) beamforming to transmit LUs' signals during the following \emph{DT} phase.
Mathematically, the ZF beamforming can be given by~\cite{LSZF,ZFBF}
\begin{equation} 
    {{\bf{W}}_{{\rm{\!ZF}}}} = \frac{{{{\bf{H}}_{{\rm{d}}}}{{\left( {{\bf{H}}_{{\rm{d}}}^H{{\bf{H}}_{{\rm{d}}}}} \right)}^{ - 1}}{\bf{P}}^{\frac{1}{2}}}}{{{{\left\| {{{\bf{H}}_{{\rm{d}}}}{{\left( {{\bf{H}}_{{\rm{d}}}^H{{\bf{H}}_{{\rm{d}}}}} \right)}^{ - 1}}} \right\|}}}}
    =\left[{\boldsymbol{w}}_{{\rm{\!Z\!F},1}},\cdots,{\boldsymbol{w}}_{{\rm{\!Z\!F},K}}\right],
    \label{LSZFBF}
\end{equation}
where ${\bf P} = {\rm{diag}}\left(p_1,\cdots,p_{\!K}\right)$ represents the power allocation matrix,
and $\left\|{\bf P}\right\|^{2} \le P_{0}$. 
It is worth noting that the optimal power allocation matrix can be computed by using the water-filling algorithm~\cite{ZFBF}.
For convenience, we further assume that $p_1 = \cdots = p_{\!K} =  {P_0}$

\underline{\emph{Active Channel Aging:}}
When ${{\bf{W}}_{{\rm{\!ZF}}}}$ has been calculated according to~\eqref{LSZFBF}, it is used by the AP in the consequent \emph{DT} phase to transmit siganls to the LUs.
In traditional MU-MISO systems, wireless channels can be assumed to remain unchanged during the channel coherence time.
However, 
in the MU-MISO system under the omnidirectional DISCO jamming attacks,
the DIOS R\&R coefficients are randomly changed whose period is about $T_{\rm P}$ and much smaller than that the length of channel coherence time $T_{\rm C}$.
As a result, ACA is introduced, and then the channel reciprocity of the wireless channels 
in traditional TDD systems is broken.
Mathematically, the time-varying channel during the $DT$ phase can be written as
\begin{equation} 
    {\bf H}_{\rm{\!D\!T}} = {\bf H}_{\rm{d}} + \!\left[ \!\!\!{\begin{array}{*{20}{c}}
        {\bf H}^{\rm t}_{\rm{D}}\!(t_{\rm{D\!T}})\\
        {\bf H}^{\rm r}_{\rm{D}}\!(t_{\rm{D\!T}})
        \end{array}} \!\!\!\right]  = {\bf H}_{\rm{d}} + \!\left[ \!\!\!{\begin{array}{*{20}{c}}
        {\bf G}^{\! H}\!{\bf \Phi}_{\rm t}\!(t_{\rm{D\!T}}){\bf H}^{\rm t}_{\rm I}\\
        {\bf G}^{\! H}\!{\bf \Phi}_{\rm r}(t_{\rm{D\!T}}){\bf H}^{\rm r}_{\rm I}
        \end{array}} \!\!\!\right],
    \label{DTChannel2}
\end{equation}
where  ${\left({\bf H}^{\rm t}_{\rm D}\!(t_{\rm{D\!T}})\!\right)^{\! H}} \!=\! \!\left[\!{{\boldsymbol{h}}^{\rm t}_{{\rm D},1}\!(t_{\rm{D\!T}})},\cdots,{{\boldsymbol{h}}^{\rm t}_{{\rm D},K_{\rm t}}\!(t_{\rm{D\!T}})}\right]^{H} \!\in\! {\mathbb{C}}^{K_{\rm t} \times {N_{\! \rm A}}}$
and ${\left({\bf H}^{\rm r}_{\rm D}\!(t_{\!P\!T})\right)^{\! H}} = \left[{{\boldsymbol{h}}^{\rm r}_{{\rm D},1}\!(t_{\!P\!T})},\cdots,{{\boldsymbol{h}}^{\rm r}_{{\rm D},K_{\rm r}}\!(t_{\!P\!T})}\right]^{H} \!\in\! {\mathbb{C}}^{K_{\rm r} \times {N_{\! \rm A}}}$
stand for the overall R\&R DIOS-jammed channels in the \emph{DT} phase, respectively.
In~\eqref{DTChannel2}, 
${\bf G }\!\in\! {\mathbb{C}}^{N_{\rm D} \times {N_{\rm A}}}$ is the channel between the AP and the DIOS, 
and ${\bf H}^{\rm t}_{\rm I} = \left[{{\boldsymbol{h}}^{\rm t}_{{\rm I},1}},\cdots,{{\boldsymbol{h}}^{\rm t}_{{\rm I},K_{\rm t}}}\right] \!\in\! {\mathbb{C}}^{N_{\rm D} \times {K_{\rm t}}}$ 
and ${\bf H}^{\rm r}_{\rm I} = \left[{{\boldsymbol{h}}^{\rm r}_{{\rm I},1}},\cdots,{{\boldsymbol{h}}^{\rm r}_{{\rm I},K_{\rm r}}}\right] \!\in\! {\mathbb{C}}^{N_{\rm D} \times {K_{\rm r}}}$ 
are the R\&R channels between the DIOS and the LUs. 

Based on~\eqref{DTChannel2}, one can see that the channel reciprocity assumption no longer holds. 
More specifically, we define the ACA channel ${\bf H}_{\rm{\!A\!C\!A}}$ as
\begin{equation} 
    {\bf H}_{\rm{\!A\!C\!A}} = {\bf H}_{\rm{\!D\!T}} - {\bf H}_{\rm{\!P\!T}} = \!\left[ \!\!\!{\begin{array}{*{20}{c}}
        {\bf H}^{\rm t}_{\rm{D}}(t_{\rm{D\!T}})\\
        {\bf H}^{\rm r}_{\rm{D}}(t_{\rm{D\!T}})
        \end{array}} \!\!\!\right].
    \label{ACAChannel}
\end{equation}

\underline{\emph{Ergodic Achievable Downlink Rate:}}
According to~\eqref{DTChannel2}, the signals received at the $k_{\rm t}$-th refractive-side LU and 
the $k_{\rm r}$-th reflective-side LU in the \emph{DT} phase ($T_{\rm P}<t_{D\!T}\le T_{\rm C}$) are expressed as~\cite{MyDIOSICC24,IOS1}
\begin{equation}
    {y^{\rm t}_{\!D\!T,k_{\rm t}}} \!= \!\!\left(\!{\boldsymbol{h}_{{\rm{d}},k_{\rm t}}^H} \!\!+ {\boldsymbol{h}_{{\rm{D}},k_{\rm t}}^H}\! \right)\! \!\sum_{u \in {\cal{K}}} {{\boldsymbol{w}_{{\rm{\!Z\!F}},u}}{x_{\!D\!T,u}}}   
      \!+\! {n_{k_{\rm t}}} ,
    \label{RXTSig}
\end{equation}
and 
\begin{equation}
    {y^{\rm r}_{\!D\!T,k_{\rm r}}} \!\!= \! \left(\!{\boldsymbol{h}_{{\rm{d}},k_{\rm r}}^H} \!\!+ {\boldsymbol{h}_{{\rm{D}},k_{\rm r}}^H} \!\right)\! \!\sum_{u \in {\cal{K}}} {{\boldsymbol{w}_{{\rm{\!Z\!F}},u}}{x_{\!D\!T,u}}}   \!+\! {n_{k_{\rm r}}} ,
    \label{RXRSig}
\end{equation}
where we assume that the transmit signals for all R\&R LUs satisfy ${\mathbb{E}}\!\left[ \left| x_{{\!D\!T},u} \right|^2 \right] = 1, u \in {\cal K}$, 
and ${{n}_{k_{\rm t}}},{{n}_{k_{\rm r}}}  \sim \mathcal{CN}\!\left(0,\delta^2\right)$ are the received AWGN. 

According to~\eqref{RXTSig} and~\eqref{RXRSig}, 
the ergodic achievable downlink rate at the  $k_{\rm t}$-th refractive-side and the $k_{\rm r}$-th reflective-side LUs
are given by
\begin{equation}
    R^{\rm t}_{k_{\rm t}}= {{\rm{log}}_{2}}\! \left( 1\!+\!\gamma^{\rm t}_{k_{\rm t}} \right),
    \label{RXTSINR}
\end{equation}
and
\begin{equation}
    R^{\rm r}_{k_{\rm r}}= {{\rm{log}}_{2}}\!\left(1\!+\!\gamma^{\rm r}_{k_{\rm r}} \right).
    \label{RXRSINR}
\end{equation}
More specifically, the SINRs of the  $k_{\rm t}$-th refractive-side and the $k_{\rm r}$-th reflective-side LUs 
 are expressed as~\cite{FLiuISAC}
\begin{equation}
    \gamma^{\rm t}_{k_{\rm t}} =  \frac{{\mathbb{E}}\!\left[\!\left|\! \left(\!{\boldsymbol{h}_{{\rm d},k_{\rm t}}} \!\!+\! {\boldsymbol{h}_{{\rm D},k_{\rm t}}^{\rm t}}\!\!(t_{D\!T})\right)^{\!H}\!{\!\boldsymbol{w}_{{\rm{\!Z\!F}},k_{\rm t}}}\!\right|^2\right]}
    { {\mathbb{E}}\!\left[{ \!\sum\limits_{u \ne {k_{\rm{t}}}}\!\! \left| \!\left( \!{\boldsymbol{h}_{{\rm d},k_{\rm t}}}\!\!+
    \!\!{\boldsymbol{h}_{{\!\rm D},k_{\rm t}}^{\rm t}}\!\!(t_{\rm{D\!T}})\!\right)^{\!H}\!\!{\!\boldsymbol{w}_{{\rm{\!Z\!F}},u}\!}\right|^2}\right] \!\!+\!\!{\delta ^2} },
    \label{RXTSINRGama}
\end{equation} 
and 
\begin{equation}
    \gamma^{\rm r}_{k_{\rm r}} = \frac{{\mathbb{E}}\!\left[\!\left|\! \left(\!{\boldsymbol{h}_{{\rm d},k_{\rm r}}} \!\!+\! {\boldsymbol{h}_{{\rm D},k_{\rm r}}^{\rm r}}\!\!(t_{D\!T})\right)^{\!H}\!{\!\boldsymbol{w}_{{\rm{\!Z\!F}},k_{\rm r}}}\!\right|^2\right]}
    {{\mathbb{E}}\!\left[{ \!\sum\limits_{u \ne {k_{\rm{r}}}}\!\! \left| \!\left( \!{\boldsymbol{h}_{{\rm d},k_{\rm r}}}\!\!+
    \!\!{\boldsymbol{h}_{{\!\rm D},k_{\rm r}}^{\rm r}}\!\!(t_{\rm{D\!T}})\!\right)^{\!H}\!\!{\!\boldsymbol{w}_{{\rm{\!Z\!F}},u}\!}\right|^2}\right] \!\!+\!\!{\delta ^2} }.
    \label{RXRSINRGama}
\end{equation} 
where the R\&R DIOS-jammed channels 
can be further given by $\left({\boldsymbol{h}_{{\rm D},k_{\rm t}}^{\rm t}}\!(t_{D\!T})\right)^{\!H} = 
{({\boldsymbol{h}_{{\rm I},k_{\rm t}}^{\rm t}})^{\! H}{{\bf{\Phi}}_{\rm t}(t_{D\!T})}{\bf G}^{\!H}}$ and 
$\left({\boldsymbol{h}_{{\rm D},k_{\rm r}}^{\rm r}}\!(t_{D\!T})\right)^{\!H} = 
{({\boldsymbol{h}_{{\rm I},k_{\rm r}}^{\rm t}})^{\! H}{{\bf{\Phi}}_{\rm r}(t_{D\!T})}{\bf G}^{\!H}}$, respectively.

From~\eqref{RXTSINR} and~\eqref{RXRSINR}, 
one can see that the DIOS-based FPJ launches omnidirectional fully-passive jamming attacks by
randomly generating the time-varying R\&R passive beamforming ${\bf{\Phi}}_{\rm t}(t_{\rm{D\!T}})$ and ${\bf{\Phi}}_{\rm r}(t_{\rm{D\!T}})$ to rapidly age wireless channels. 
The fully-passive jamming is also referred to as active channel aging interference (ACAI)~\cite{MyDRIS5}.
Consequently, the ergodic achievable sum rate is given by ${R_{{\rm{sum}}}} = {R^{\rm t}_{{\rm{sum}}}} + {R^{\rm r}_{{\rm{sum}}}} = \sum\nolimits_{{k_{\rm{t}}} \in {\cal K}_{\rm t}}\! { {R_{{k_{\rm{t}}}}^{\rm{t}}}} + 
\sum\nolimits_{{k_{\rm{r}}} \in {\cal K}_{\rm r}}\! { {R_{{k_{\rm{r}}}}^{\rm{r}}}}$.

\subsection{Channel Model}\label{ChannelModel}
In the MU-MISO system attacked by the DIOS-based FPJ, the AP-DIOS channel $\bf G$ in~\eqref{DTChannel2} 
is constructed based on the near-field model because the DIOS is implemented near to the AP.
Mathematically, $\bf G$ in~\eqref{DTChannel2} is is modeled as~\cite{DIOSNearFarModel,DIOSNearFarModel1}
\begin{equation}
    \begin{split}
    {\bf G} \!= \! {\sqrt{\!\!{\mathscr{L}}_{\rm{\!G}}}} \! 
    \left(\! \sqrt {\frac{{{\varepsilon_{\rm{\!G}}}}}{{1\!+\!{\varepsilon_{\rm{\!G}}}}}}\!{{\widehat{\bf{G}}}^{\!{\rm{LOS}}}}  \!+\! 
    \sqrt {\frac{1}{{1\!+\!{\varepsilon_{\rm{\!G}}}}}}\!{{\widehat {\bf{G}}}^{{\rm{NLOS}}}} \! \right),
    \end{split}
\label{Ricianchan}
\end{equation}
where ${\mathscr{L}}_{\rm{\!G}}$ is the large-scale channel fading of ${\bf G}$  
and ${{\varepsilon_{\rm{\!G}}}}$ represents the Rician factor for ${\bf G}$.  
In~\eqref{Ricianchan},
${{\widehat {\bf{G}}}^{{\rm{NLOS}}}}$ follows Rayleigh fading~\cite{DIOSNearFarModel,BookFarFeild}, i.e.,
the elements $\left[ {{\widehat {\bf{G}}}^{{\rm{NLOS}}}} \right]_{n,s} \sim \mathcal{CN}\left(0,1\right), n=1, \cdots, N_{\rm A}, s=1, \cdots, N_{\rm D}$.
Furthermore, the elements of ${{\widehat {\bf{G}}}^{{\rm{LOS}}}}$ are given by~\cite{MyDRIS5,DIOSNearFarModel1}
\begin{equation}
    \left[{{\widehat {\bf{G}}}^{{\rm{LOS}}}}\right]_{n,s} \!\! = \! {e^{ \!- j\frac{{2\pi }}{\lambda }\left( {{d_{n,s}} - {d_n}} \right)}},
    \label{GLOS}
\end{equation}
where $\lambda$ is the wavelength of transmit signals, and
${d_{n,s}}$ and ${d_n}$ are the distance between the $n$-th ULA antenna and the $s$-th DIOS element
and the distance between the $n$-th ULA antenna and the origin of the DIOS, respectively. 

Moreover, the R\&R DIOS-LU channels ${\bf H}^{\rm t}_{\!{\rm{I}}}$ and ${\bf H}^{\rm r}_{\!{\rm{I}}}$, 
and the direct AP-LU channel ${\bf H}_{\!{\rm{d}}}$~\eqref{DTChannel2}
are modeled based on the far-field model~\cite{BookFarFeild}:
\begin{alignat}{1}
    &{{\bf{H}}^{\rm t}_{\rm{I}}}  
     \!=\! \left[\! {{\sqrt{\!{\!{\mathscr{L}}^{\rm t}_{{\rm{I}},1}}}}{{\widehat {\boldsymbol{h}}}^{\rm t}_{{\rm{I}},1}},  \!\cdots , 
     \!{\sqrt{\!{\!{\mathscr{L}}^{\rm t}_{{\rm{I}},K_{\rm t}}}}}{{\widehat {\boldsymbol{h}}}^{\rm t}_{{\rm{I}},K_{\rm t}}}} \right], \label{HIkeq}\\
     &{{\bf{H}}^{\rm r}_{\rm{I}}}  
     \!=\! \left[\! {{\sqrt{\!{\!{\mathscr{L}}^{\rm r}_{{\rm{I}},1}}}}{{\widehat {\boldsymbol{h}}}^{\rm r}_{{\rm{I}},1}},  \!\cdots , 
     \!{\sqrt{\!{\!{\mathscr{L}}^{\rm r}_{{\rm{I}},K_{\rm r}}}}}{{\widehat {\boldsymbol{h}}}^{\rm r}_{{\rm{I}},K_{\rm r}}}} \right], \label{HIkeqR}\\
    &{{\bf{H}}_{\rm{d}}}  
    \!=\! \left[\! {{\sqrt{\!{\!{\mathscr{L}}_{{\rm{d}},1}}}}{{\widehat {\boldsymbol{h}}}_{{\rm{d}},1}}, 
    \! \cdots, \! {\sqrt{\!{\!{\mathscr{L}}_{{\rm{d}},K}}}}{{\widehat {\boldsymbol{h}}}_{{\rm{d}},K}}} \right],
    \label{Hdkeq}
\end{alignat}
where ${{\mathscr{L}}^{\rm t}_{{\rm{I}},k_{\rm t}}}, k_{\rm t} \in {\cal K}_{\rm t}$, ${{\mathscr{L}}^{\rm r}_{{\rm{I}},k_{\rm r}}}, k_{\rm r} \in {\cal K}_{\rm r}$,
and ${{\mathscr{L}}_{{\rm{d}},k}}, k \in {\cal K}$ denote the large-scale channel fading coefficients. 
The $n$-th elements in ${{\widehat {\boldsymbol{h}}}^{\rm t}_{{\rm{I}},k_{\rm t}}}$, ${{\widehat {\boldsymbol{h}}}^{\rm r}_{{\rm{I}},k_{\rm r}}}$,
and ${{\widehat {\boldsymbol{h}}}_{{\rm{d}},k}}$ are modeled as independent and identically distributed (i.i.d.) Gaussian random variables
with mean zero and variance 1, 
and are assumed to be independent over $n$, $n=1,\cdots,N\!_{\rm A}$~\cite{ChannelElementMode}.

\section{Ergodic Achievable Downlink Rate Under DIOS-Based Fully-Passive Jamming Attacks}\label{DIOSFPJImpact}
In this section, we derive the statistical characteristics of the DIOS-jammed channel to characterize the jamming impact of the DIOS-based FPJ
for two IOS models, i.e., the constant-amplitude model and the variable-amplitude model in Section~\ref{CMDMDIOS}.
In Section~\ref{DISCORate}, we further derive a lower bound of the ergodic achievable downlink rate
based on the statistical caracteristics of the DIOS-jammed channel.

\subsection{Statistical Characteristics of Active Channel Aging}\label{CMDMDIOS}
According to~\eqref{RXTSINR} and~\eqref{RXRSINR}, the time-varying DIOS R\&R coefficients fail channel reciprocity.
As a result,  the ACAI is introduced to launch omnidirectional fully-passive jamming attacks. 
Therefore, the impact of the DIOS-based FPJ is directly dependent on the characteristics of the time-varying R\&R DIOS-jammed channels ${\bf{H}}^{\rm t}_{\rm {D}}(t)$ and ${\bf{H}}^{\rm r}_{\rm {D}}(t)$.
We first assume that the DIOS R\&R coefficients are constant, i.e., ${\alpha}^{\rm t}_{s}(t) = {\alpha}^{\rm r}_{s}(t) = \frac{\sqrt{2}}{2}, \forall s$.
Then, the statistical characteristics of ${\bf{H}}^{\rm t}_{\rm {D}}(t)$ and ${\bf{H}}^{\rm r}_{\rm {D}}(t)$ are given in Proposition~\ref{Proposition1}.

\newtheorem{proposition}{Proposition}
\begin{proposition}
    \label{Proposition1}
    For a constant-amplitude DIOS, the i.i.d. elements in ${\bf{H}}^{\rm t}_{\rm {D}}(t)$ and ${\bf{H}}^{\rm r}_{\rm {D}}(t)$ converge in distribution to 
    $\mathcal{CN}\!\left( {0, {{{{\mathscr{L}}\!_{{\rm G}}}{{\mathscr{L}}^{\rm t}_{{\rm I},k_{\rm t}}}{N\!_{\rm D}}}}/{2} } \right)$ and 
    $\mathcal{CN}\!\left( {0, {{{{\mathscr{L}}\!_{{\rm G}}}{{\mathscr{L}}^{\rm r}_{{\rm I},k_{\rm r}}}{N\!_{\rm D}}}}/{2} } \right)$
    as $N_{\rm D} \to \infty$, i.e.,
    \begin{equation}
        \left[{\bf{H}}^{\rm t}_{\rm {D}}(t)\right]_{n,k_{\rm t}} \mathop  \to \limits^{\rm{d}}
        \mathcal{CN}\!\!\left( {0, \frac{{{{\mathscr{L}}\!_{{\rm G}}}{{\mathscr{L}}^{\rm t}_{{\rm I},k_{\rm t}}}{N\!_{\rm D}}}}{2} } \right),
        \label{DIOSChaT}
    \end{equation}
    and
    \begin{equation}
        \left[{\bf{H}}^{\rm r}_{\rm {D}}(t)\right]_{n,k_{\rm r}} \mathop  \to \limits^{\rm{d}}
        \mathcal{CN}\!\!\left( {0, \frac{{{{\mathscr{L}}\!_{{\rm G}}}{{\mathscr{L}}^{\rm r}_{{\rm I},k_{\rm r}}}{N\!_{\rm D}}}}{2} } \right),
        \label{DIOSChaR}
    \end{equation}
\end{proposition}
where $0 \le t \le T_{\rm C}$, $n = 1,\cdots, N_{\rm A}$, $k_{\rm t} \in {\cal K}_{\rm t}$, and $k_{\rm r} \in {\cal K}_{\rm r}$.

\begin{IEEEproof}
    See Appendix~\ref{AppendixA}.
\end{IEEEproof}

It should be noted that, in practice, the DIOS must have a large number of elements to handle the multiplicative large-scale channel fading in the DIOS-jammed R\&R channels.
From Proposition~\ref{Proposition1}, a property of the DIOS-based FPJ implemented by a constant-amplitude IOS  is that
its jamming impact 
does not depend on nor the number of its phase quantization bits nor the stochastic distribution of the DIOS R\&R phase shifts.
Namely, we can use a one-bit quantization IOS whose R\&R phase shifts follow the simple uniform distribution 
to effectively implement the DIOS-based PFJ.
 
For the variable-amplitude IOS model built in Section~\ref{DIOSFPJ}, 
we denote the probability of the R\&R phase shift $\varphi^{\rm t}_{s}$ and $\varphi^{\rm r}_{s}$ taking the $m$-th value in 
$\Theta$, i.e., $\theta_{m}$ as $P_{m}, \forall s$.
As a result, the statistical characteristics in Proposition~\ref{Proposition1} shift to Proposition~\ref{Proposition2}.
\begin{proposition}
    \label{Proposition2}
    For a variable-amplitude DIOS, the i.i.d. elements in ${\bf{H}}^{\rm t}_{\rm {D}}(t)$ and ${\bf{H}}^{\rm r}_{\rm {D}}(t)$ converge in distribution to 
    $\mathcal{CN}\!\left( {0, {{{{\mathscr{L}}\!_{{\rm G}}}{{\mathscr{L}}^{\rm t}_{{\rm I},k_{\rm t}}}{N\!_{\rm D}}{\mu}}} } \right)$ and 
    $\mathcal{CN}\!\left( {0, {{{{\mathscr{L}}\!_{{\rm G}}}{{\mathscr{L}}^{\rm r}_{{\rm I},k_{\rm r}}}{N\!_{\rm D}}({1-\mu})}} } \right)$
    as $N_{\rm D} \to \infty$, i.e.,
    \begin{equation}
        \left[{\bf{H}}^{\rm t}_{\rm {D}}(t)\right]_{n,k_{\rm t}} \mathop  \to \limits^{\rm{d}}
        \mathcal{CN}\!\!\left( {0, {{{\mathscr{L}}\!_{{\rm G}}}{{\mathscr{L}}^{\rm t}_{{\rm I},k_{\rm t}}}{N\!_{\rm D}}{\mu}} } \right),
        \label{DIOSChaT1}
    \end{equation}
    and
    \begin{equation}
        \left[{\bf{H}}^{\rm r}_{\rm {D}}(t)\right]_{n,k_{\rm r}} \mathop  \to \limits^{\rm{d}}
        \mathcal{CN}\!\!\left( {0, {{{\mathscr{L}}\!_{{\rm G}}}{{\mathscr{L}}^{\rm r}_{{\rm I},k_{\rm r}}}{N\!_{\rm D}}\left({1-\mu}\right)} } \right),
        \label{DIOSChaR1}
    \end{equation}
\end{proposition}
where $\mu = \sum\nolimits_{m = 1}^{{2^b}} {{P_m}{{\left( {\xi _m^{\rm{t}}} \right)}^2}} = 1 - \left(\sum\nolimits_{m = 1}^{{2^b}} {{P_m}{{\left( {\xi _m^{\rm{r}}} \right)}^2}}\right)$.

\begin{IEEEproof}
    See Appendix~\ref{AppendixB}.
\end{IEEEproof}

Based on Proposition~\ref{Proposition2}, we can see that the jamming impact of the DIOS-based FPJ implemented by an IOS with variable amplitudes
depends on  the number of DIOS phase quantization bits and the stochastic distribution of the DIOS R\&R phase shifts.
Furthermore, the impact on the refractive-side LUs and that on the reflective-side LUs
are mutually exclusive, and the trade-off between them can be tuned by stochastic distribution of the DIOS R\&R phase shifts.

\subsection{Lower Bound of Ergodic Achievable Downlink Rate}\label{DISCORate}
In this section, we aim to quantify the impact of the DIOS-based FPJ on LUs.
The ergodic achievable R\&R sum rate $R^{\rm t}_{\rm{sum}}$ and $R^{\rm r}_{\rm{sum}}$ of the refractive-side users and the reflective-side users have been given based on 
the definitions in~\eqref{RXTSINR} and~\eqref{RXRSINR}.
According to~\eqref{RXTSINRGama} and~\eqref{RXRSINRGama}, 
$R^{\rm t}_{\rm{sum}}$ and $R^{\rm r}_{\rm{sum}}$ 
can be expressed as 
\begin{alignat}{1}
    \nonumber
    R_{{\rm{sum}}}^{\rm t} &= \!\sum\limits_{{k_{\rm t}} \in {{\cal K}_t}}\!\! {R^{\rm t}_{k_{\rm t}}} 
    = \!\sum\limits_{{k_{\rm t}} \in {{\cal K}_{\rm t}}}\! {{\rm{log}}_{2}}\! \left( 1\!+\!\gamma^{\rm t}_{k_{\rm t}} \right) \\
    & = \!\!\!\sum\limits_{{k_{\rm t}} \in {{\cal K}_{\rm t}}}\!\!{{\rm{log}}_{2}}\!\!\left(\!\! 1\!+\! 
    \frac{{\mathbb{E}}\!\left[\!\left|\! \left(\!{\boldsymbol{h}_{{\rm d},k_{\rm t}}} \!\!+\! 
    {\boldsymbol{h}_{{\rm D},k_{\rm t}}^{\rm t}}\!\!(t_{D\!T})\right)^{\!H}\!{\!\boldsymbol{w}_{{\rm{\!Z\!F}},k_{\rm t}}}\!\right|^2\right]}
    { {\mathbb{E}}\!\!\left[{ \!\sum\limits_{u \ne {k_{\rm{t}}}}\!\! \left| \!\left( \!{\boldsymbol{h}_{{\rm d},k_{\rm t}}}\!\!+
    \!\!{\boldsymbol{h}_{{\!\rm D},k_{\rm t}}^{\rm t}}\!\!(t_{\rm{D\!T}})\!\right)^{\!H}\!\!{\!\boldsymbol{w}_{{\rm{\!Z\!F}},u}\!}\right|^2}\right] \!\!+\!\!{\delta ^2} }\!\!\right), 
   \label{JensenInet}
\end{alignat}
and
\begin{alignat}{1}
    \nonumber
    R_{{\rm{sum}}}^{\rm r} &= \!\sum\limits_{{k_{\rm r}} \in {{\cal K}_r}}\!\!{R^{\rm r}_{k_{\rm r}}} 
    = \!\sum\limits_{{k_{\rm r}} \in {{\cal K}_{\rm r}}}\! {{\rm{log}}_{2}}\! \left( 1\!+\!\gamma^{\rm r}_{k_{\rm r}} \right) \\
    & = \!\!\!\sum\limits_{{k_{\rm r}} \in {{\cal K}_{\rm r}}}\!\!{{\rm{log}}_{2}}\!\!\left(\!\! 1\!+\! 
    \frac{{\mathbb{E}}\!\left[\!\left|\! \left(\!{\boldsymbol{h}_{{\rm d},k_{\rm r}}} \!\!+\! 
    {\boldsymbol{h}_{{\rm D},k_{\rm r}}^{\rm r}}\!\!(t_{D\!T})\right)^{\!H}\!{\!\boldsymbol{w}_{{\rm{\!Z\!F}},k_{\rm r}}}\!\right|^2\right]}
    { {\mathbb{E}}\!\!\left[{ \!\sum\limits_{u \ne {k_{\rm{r}}}}\!\! \left| \!\left( \!{\boldsymbol{h}_{{\rm d},k_{\rm r}}}\!\!+
    \!\!{\boldsymbol{h}_{{\!\rm D},k_{\rm r}}^{\rm r}}\!\!(t_{\rm{D\!T}})\!\right)^{\!H}\!\!{\!\boldsymbol{w}_{{\rm{\!Z\!F}},u}\!}\right|^2}\right] \!\!+\!\!{\delta ^2} }\!\!\right), 
   \label{JensenIner} 
\end{alignat}

However, the achievable downlink rates given in~\eqref{JensenInet} and~\eqref{JensenIner} are implicit.
To this end, more-explicit lower bounds of $R^{\rm t}_{\rm{sum}}$ and $R^{\rm r}_{\rm{sum}}$ are required.
Therefore, we future derive the more useful lower bounds in the following Theorem~\ref{Theorem1} and Theorem~\ref{Theorem2}.
More specifically, the more-explicit lower bounds of $R^{\rm t}_{\rm{sum}}$ and $R^{\rm r}_{\rm{sum}}$ in Theorem~\ref{Theorem1}
are derived based on the constant-amplitude IOS model. 
\newtheorem{theorem}{Theorem}
\begin{theorem}
\label{Theorem1}
For a constant-amplitude DIOS,
the lower bound on the ergodic achievable R\&R sum rate $R^{\rm t}_{\rm{sum}}$ and $R^{\rm r}_{\rm{sum}}$
converges in probability towards a fixed value as $N_{\rm D} \to \infty $, i.e., 
\begin{alignat}{1}
    R_{{\rm{sum}}}^{\rm t} & {\mathop  \to \limits^{\rm{p}} }  
    \!\sum\limits_{{k_{\rm t}} \in {{\cal K}_{\rm t}}}\!\!{\rm{log}_2}\!\!\left(\!\!1\!+\!\!
    { \frac{ {\mathbb{E}}\!\!\left[\!\frac{P_0}{{\rm {tr}}{{\left( \!{{\bf{H}}_{\rm{d}}^H{{\bf{H}}_{\rm{d}}}} \right)}^{ \!- 1}}}\right] \!+ 
    \!\frac{P_0{{\mathscr{L}}\!_{{\rm G}}}{{\mathscr{L}}_{{\rm I},k_{\rm t}}^{\rm t}}N\!_{\rm D}}{2K}}
    { \frac{P_0{{{\mathscr{L}}\!_{{\rm G}}}\!\left(\!\sum\nolimits_{u \ne {k_{\rm{t}}}}\!{{\mathscr{L}}}_{{\rm I},u}\!\right)}{N\!_{\rm D}}}{2K} \!+\!\delta^2}}\!\right) \label{JensenInetLBStaCha}\\
    & \ge \!\sum\limits_{{k_{\rm t}} \in {{\cal K}_{\rm t}}}\!\!{\rm{log}_2}\!\!\left(\!\!1\!+\!\!
    { \frac{ \frac{2P_0K(\!N\!_{\rm A}-K)}{ \!{\sum\nolimits_{k = 1}^K \!{\frac{1}{ {{\mathscr{L}}}_{{\rm d},k} }}}  } \!+ 
    \!{P_0{{\mathscr{L}}\!_{{\rm G}}}{{\mathscr{L}}_{{\rm I},k_{\rm t}}^{\rm t}}N\!_{\rm D}}}
    {{P_0{{{\mathscr{L}}\!_{{\rm G}}}\!\left(\!\sum\nolimits_{u \ne {k_{\rm{t}}}}\!{{\mathscr{L}}}_{{\rm I},u}\!\right)}{N\!_{\rm D}}} \!+\!2K\delta^2}}\!\right)
\label{JensenInetLB}
\end{alignat}
and 
\begin{alignat}{1}
    R_{{\rm{sum}}}^{\rm r} & {\mathop  \to \limits^{\rm{p}} }  
    \!\sum\limits_{{k_{\rm r}} \in {{\cal K}_{\rm r}}}\!\! {\rm{log}_2}\!\!\left(\!\!1\!+\!\!
    { \frac{ {\mathbb{E}}\!\!\left[\!\frac{P_0}{{\rm {tr}}{{\left( \!{{\bf{H}}_{\rm{d}}^H{{\bf{H}}_{\rm{d}}}} \right)}^{ \!- 1}}}\right] \!+ 
    \!\frac{P_0{{\mathscr{L}}\!_{{\rm G}}}{{\mathscr{L}}_{{\rm I},k_{\rm r}}^{\rm r}}N\!_{\rm D}}{2K}}
    { \frac{P_0{{{\mathscr{L}}\!_{{\rm G}}}\!\left(\!\sum\nolimits_{u \ne {k_{\rm{r}}}}\!{{\mathscr{L}}}_{{\rm I},u}\!\right)}{N\!_{\rm D}} }{2K} \!+\!\delta^2}}\!\right) \label{JensenInerLBStaCha}\\
    & \ge \!\sum\limits_{{k_{\rm r}} \in {{\cal K}_{\rm r}}}\!\!{\rm{log}_2}\!\!\left(\!\!1\!+\!\!
    { \frac{ \frac{2KP_0(\!N\!_{\rm A}-K)}{ \!{\sum\nolimits_{k = 1}^K \!{\frac{1}{ {{\mathscr{L}}}_{{\rm d},k} }}}  } \!+ 
    \!{P_0{{\mathscr{L}}\!_{{\rm G}}}{{\mathscr{L}}_{{\rm I},k_{\rm r}}^{\rm r}}N\!_{\rm D} }}
    {{P_0{{{\mathscr{L}}\!_{{\rm G}}}\!\left(\!\sum\nolimits_{u \ne {k_{\rm{r}}}}\!{{\mathscr{L}}}_{{\rm I},u}\!\right)}{N\!_{\rm D}} } \!+\!2K\delta^2}}\!\right)
\label{JensenInerLB}
\end{alignat}
\end{theorem}

\begin{IEEEproof}
Conditioned on the fact that the random variables ${\boldsymbol{h}_{{\rm d},k_{\rm t}}}$,
${\boldsymbol{h}_{{\rm D},k_{\rm t}}^{\rm t}}\!\!(t_{D\!T})$,
and ${\boldsymbol{w}_{{\rm{\!Z\!F}},k_{\rm t}}}$ are independent of each other, 
we can reduce numerator terms in~\eqref{JensenInet} and~\eqref{JensenIner}
to the following forms:
\begin{alignat}{1}
    \nonumber
    &{{\mathbb{E}}\!\left[\!\left|\!\left({\boldsymbol{h}_{{\rm d},k_{\rm t}}} \!\!+\! 
    {\boldsymbol{h}_{{\rm D},k_{\rm t}}^{\rm t}}\!\!(t_{D\!T})\right)^{\!H}\!{\!\boldsymbol{w}_{{\rm{\!Z\!F}},k_{\rm t}}}\!\right|^2\right]} \\
    & =\! {\mathbb{E}}\!\!\left[\left|\! {\boldsymbol{h}^H_{{\rm d},k_{\rm t}}}\!{\!\boldsymbol{w}_{{\rm{\!Z\!F}},k_{\rm t}}}\!\right|^2\right] \!+\!
    {{\mathbb{E}}\!\left[\!\left| \!\left({\boldsymbol{h}_{{\rm D},k_{\rm t}}^{\rm t}}\!\!(t_{D\!T})\!\right)^H\!{\!\boldsymbol{w}_{{\rm{\!Z\!F}},k_{\rm t}}}\!\right|^2\right]} \label{ExOmatPre}\\
    & =  {\mathbb{E}}\!\!\left[{\frac{P_0}{\left\| {{\bf{H}}_{\rm{d}}{{\left( {{\bf{H}}_{\rm{d}}^H{{\bf{H}}_{\rm{d}}}} \right)}^{ - 1}}} \right\|^2}}\right] \!+\!
    {{\mathbb{E}}\!\left[\!\left| \!\left({\boldsymbol{h}_{{\rm D},k_{\rm t}}^{\rm t}}\!\!(t_{D\!T})\!\right)^H\!{\!\boldsymbol{w}_{{\rm{\!Z\!F}},k_{\rm t}}}\!\right|^2\right]} ,
    \label{ExOmat}
\end{alignat}
and 
\begin{alignat}{1}
    \nonumber
        &{{\mathbb{E}}\!\left[\!\left|\! \left({\boldsymbol{h}_{{\rm d},k_{\rm r}}} \!\!+\! 
        {\boldsymbol{h}_{{\rm D},k_{\rm r}}^{\rm r}}\!\!(t_{D\!T})\right)^{\!H}\!{\!\boldsymbol{w}_{{\rm{\!Z\!F}},k_{\rm r}}}\!\right|^2\right]} \\
        &=\! {\mathbb{E}}\!\!\left[\left| {\boldsymbol{h}^H_{{\rm d},k_{\rm r}}}\!{\!\boldsymbol{w}_{{\rm{\!Z\!F}},k_{\rm r}}}\!\right|^2\right] \!+\!
        {{\mathbb{E}}\!\left[\!\left|\left({\boldsymbol{h}_{{\rm D},k_{\rm r}}^{\rm r}}\!\!(t_{D\!T})\!\right)^H\!{\!\boldsymbol{w}_{{\rm{\!Z\!F}},k_{\rm r}}}\!\right|^2\right]}     \label{ExOmarPre} \\
        &=  {\mathbb{E}}\!\!\left[{\frac{P_0}{\left\| {{\bf{H}}_{\rm{d}}{{\left( {{\bf{H}}_{\rm{d}}^H{{\bf{H}}_{\rm{d}}}} \right)}^{ - 1}}} \right\|^2}}\right] \!+\!
        {{\mathbb{E}}\!\left[\!\left|\left({\boldsymbol{h}_{{\rm D},k_{\rm r}}^{\rm r}}\!\!(t_{D\!T})\!\right)^H\!{\!\boldsymbol{w}_{{\rm{\!Z\!F}},k_{\rm r}}}\!\right|^2\right]}.
    \label{ExOmar}
\end{alignat}

In~\eqref{ExOmat} and~\eqref{ExOmar}, the expectations ${{\mathbb{E}}\!\!\left[\!\left| \!\left({\boldsymbol{h}_{{\rm D},k_{\rm t}}^{\rm t}}\!\!(t_{D\!T})\!\right)^H\!{\!\boldsymbol{w}_{{\rm{\!Z\!F}},k_{\rm t}}}\!\right|^2\right]}$
and ${{\mathbb{E}}\!\!\left[\!\left|\left({\boldsymbol{h}_{{\rm D},k_{\rm r}}^{\rm r}}\!\!(t_{D\!T})\!\right)^H\!{\!\boldsymbol{w}_{{\rm{\!Z\!F}},k_{\rm r}}}\!\right|^2\right]}$ 
can be reduced to 
\begin{alignat}{1}
    \nonumber
    &{{\mathbb{E}}\!\left[\!\left| \!\left({\boldsymbol{h}_{{\rm D},k_{\rm t}}^{\rm t}}\!\!(t_{D\!T})\!\right)^H\!{\!\boldsymbol{w}_{{\rm{\!Z\!F}},k_{\rm t}}}\!\right|^2\right]} \\
    &\;\;\;\;\;\;\;\;= \!\sum\limits_{n = 1}^{{N_{\rm{A}}}} \!{\mathbb{E}}\!\!\left[\!{\left|\!{\left[{\boldsymbol{h}_{{\rm D},k_{\rm t}}^{\rm t}}\!\!(t_{D\!T})\right]_n}\right|^2}\right]\!
    {\mathbb{E}}\!\!\left[{\!\left|{\left[ \boldsymbol{w}_{{\rm{\!Z\!F}},k_{\rm t}}\right]_n}\right|^2}\!\right],
      \label{HDeleWkt}
\end{alignat}
and
\begin{alignat}{1}
    \nonumber
    &{{\mathbb{E}}\!\left[\!\left| \!\left({\boldsymbol{h}_{{\rm D},k_{\rm R}}^{\rm r}}\!\!(t_{D\!T})\!\right)^H\!{\!\boldsymbol{w}_{{\rm{\!Z\!F}},k_{\rm r}}}\!\right|^2\right]} \\
    &\;\;\;\;\;\;\;\;=  \!\sum\limits_{n = 1}^{{N_{\rm{A}}}} \!{\mathbb{E}}\!\!\left[\!{\left|\!{\left[{\boldsymbol{h}_{{\rm D},k_{\rm r}}^{\rm r}}\!\!(t_{D\!T})\right]_n}\right|^2\!}\right]\!
    {\mathbb{E}}\!\!\left[{
    \left|{\left[\boldsymbol{w}_{{\rm{\!Z\!F}},k_{\rm r}}\right]_n}\right|^2}\!\right],
      \label{HDeleWkr}
    \end{alignat}
where ${\left[{\boldsymbol{h}_{{\rm D},k_{\rm t}}^{\rm t}}\!\!(t_{D\!T})\right]_n}$, ${\left[\boldsymbol{w}_{{\rm{\!Z\!F}},k_{\rm t}}\right]_n}$, 
${\left[{\boldsymbol{h}_{{\rm D},k_{\rm r}}^{\rm r}}\!\!(t_{D\!T})\right]_n}$, and 
${\left[\boldsymbol{w}_{{\rm{\!Z\!F}},k_{\rm r}}\right]_n}$ represent the $n$-th variables of ${\boldsymbol{h}_{{\rm D},k_{\rm t}}^{\rm t}}\!\!(t_{D\!T})$,
$\boldsymbol{w}_{{\rm{\!Z\!F}},k_{\rm t}}$, ${\boldsymbol{h}_{{\rm D},k_{\rm r}}^{\rm r}}\!\!(t_{D\!T})$, and 
$\boldsymbol{w}_{{\rm{\!Z\!F}},k_{\rm r}}$, respectively.

Based on the statistical characteristics derived in Proposition~\ref{Proposition1}, we have
\begin{alignat}{1}
    {{\mathbb{E}}\!\left[\!\left| \!\left({\boldsymbol{h}_{{\rm D},k_{\rm t}}^{\rm t}}\!\!(t_{D\!T})\!\right)^H\!{\!\boldsymbol{w}_{{\rm{\!Z\!F}},k_{\rm t}}}\!\right|^2\right]} 
    & {\mathop  \to \limits^{\rm{p}} } 
    \frac{{{\mathscr{L}}\!_{{\rm G}}}{{\mathscr{L}}_{{\rm I},k_{\rm t}}^{\rm t}}N\!_{\rm D} {{\mathbb{E}}\!\left[\!\left\| \!{\boldsymbol{w}_{{\rm{\!Z\!F}},k_{\rm t}}}\!\right\|^2\right]} }{2} \label{Thdktxx} \\
    & {\mathop  \to \limits^{\rm{p}} }  \frac{{{\mathscr{L}}\!_{{\rm G}}}{{\mathscr{L}}_{{\rm I},k_{\rm t}}^{\rm t}}N\!_{\rm D} P_0 }{2K},
    \label{Thdkt}
\end{alignat}
and
\begin{alignat}{1}
    {{\mathbb{E}}\!\left[\!\left| \!\left({\boldsymbol{h}_{{\rm D},k_{\rm R}}^{\rm r}}\!\!(t_{D\!T})\!\right)^H\!{\!\boldsymbol{w}_{{\rm{\!Z\!F}},k_{\rm r}}}\!\right|^2\right]} &
    {\mathop  \to \limits^{\rm{p}} } 
    \frac{{{\mathscr{L}}\!_{{\rm G}}}{{\mathscr{L}}_{{\rm I},k_{\rm r}}^{\rm r}}N\!_{\rm D} {{\mathbb{E}}\!\left[\!\left\| \!{\boldsymbol{w}_{{\rm{\!Z\!F}},k_{\rm r}}}\!\right\|^2\right]}}{2} \label{Thdkrxx}\\
    &{\mathop  \to \limits^{\rm{p}} }\frac{{{\mathscr{L}}\!_{{\rm G}}}{{\mathscr{L}}_{{\rm I},k_{\rm r}}^{\rm r}}N\!_{\rm D} P_0}{2K},
    \label{Thdkr}
\end{alignat}
as ${N_{\rm{D}}} \to \infty$.

Moreover, we can reduce the term in~\eqref{ExOmat} and~\eqref{ExOmar} based on the Jensen inequality, i.e.,
\begin{equation}
    {\mathbb{E}}\!\!\left[\!\frac{{1}}{{\left\| {{\bf{H}}_{\rm{d}}{{\left( {{\bf{H}}_{\rm{d}}^H{{\bf{H}}_{\rm{d}}}} \right)}^{ - 1}}} \right\|^2}}\!\right] \ge 
    \frac{{1}}{{\mathbb{E}}\!\!\left[{\!\left\| {{\bf{H}}_{\rm{d}}{{\left( {{\bf{H}}_{\rm{d}}^H{{\bf{H}}_{\rm{d}}}} \right)}^{ - 1}}} \right\|^2}\!\right]}.
    \label{JensIne}
\end{equation}
Using an idiomatic trick that ${\rm {tr}}\!\left({\bf A}^{\!H}\!{\bf A}\right) = \left\|{\bf A}\right\|^2$, we can obtain that
\begin{equation}
    {{\mathbb{E}}\!\!\left[{\!\left\| {{\bf{H}}_{\rm{d}}{{\left( {{\bf{H}}_{\rm{d}}^H{{\bf{H}}_{\rm{d}}}} \right)}^{ - 1}}} \right\|^2}\!\right]} = 
    {\mathbb{E}}\!\left[\!{\rm {tr}}\!\left({{\!\left( {{\bf{H}}_{\rm{d}}^H{{\bf{H}}_{\rm{d}}}} \right)}^{ \!- 1}}\right)\!\right].
    \label{TrNorm}
\end{equation}

Based on the channel model of ${\bf H}_{\rm d}$ in~\eqref{Hdkeq}, we further reduce~\eqref{TrNorm} to 
\begin{equation}
    {\mathbb{E}}\!\left[\!{\rm {tr}}\!\left({{\!\left( {{\bf{H}}_{\rm{d}}^H{{\bf{H}}_{\rm{d}}}} \right)}^{ \!- 1}}\right)\!\right] =
    {\mathbb{E}}\!\left[\!{\rm {tr}}\!\left({ \!{\bf \Lambda}^{-1}\!\left( \!{ \widehat {\bf{H}}_{\rm{d}}^{\!H}\!{\widehat {\bf{H}}_{\rm{d}}}} \right)^{ \!- 1}}\right)\!\right],
    \label{TrNormRedu}
\end{equation}
where ${\bf \Lambda} = {\rm{diag}}\!\left({{\mathscr{L}}_{{\rm d},1}},{{\mathscr{L}}_{{\rm d},2}},\cdots,{{\mathscr{L}}_{{\rm d},K}}\right)$ 
and ${\widehat {\bf{H}}}_{\rm d} = \left[{\widehat{\boldsymbol{h}}_{{\rm d},1}},{\widehat{\boldsymbol{h}}_{{\rm d},2}},
\cdots,{\widehat{\boldsymbol{h}}_{{\rm d},K}}  \right]$.
Consequently, ${ \widehat {\bf{H}}_{\rm{d}} ^{\!H}\!{\widehat {\bf{H}}_{\rm{d}}}}$ is a central complex Wishart matrix.

Exploiting the property of complex Wishart matrices~\cite{Wishart}, we can further reduce~\eqref{TrNorm} to the following form:
\begin{equation} 
    {\mathbb{E}}\!\left[\!{\rm {tr}}{{\!\left( {{\bf{H}}_{\rm{d}}^H{{\bf{H}}_{\rm{d}}}} \right)}^{ \!- 1}}\right] = 
    \frac{1}{{{N_{\rm{A}}} - K}}\sum\limits_{k = 1}^K \frac{1}{{{\mathscr{L}}\!_{{\rm d},k}}}.
    \label{TrNormSt}
\end{equation}
As a result, we can obtain the following inequalities:
\begin{alignat}{1}
    \nonumber
    &{{\mathbb{E}}\!\left[\!\left| \left({\boldsymbol{h}_{{\rm d},k_{\rm t}}} \!\!+\! 
    {\boldsymbol{h}_{{\rm D},k_{\rm t}}^{\rm t}}\!\!(t_{D\!T})\right)^{\!H}\!{\!\boldsymbol{w}_{{\rm{\!Z\!F}},k_{\rm t}}}\!\right|^2\right]}  \ge \\
    &\;\;\;\;\;\;\;\;\;\;\;\;\; { \frac{P_0(\!N\!_{\rm A}-K)}{ {\sum\limits_{k = 1}^K \!{\frac{1}{ {{\mathscr{L}}}_{{\rm d},k} }}}  } \!+ 
    \!\frac{{P_0{{\mathscr{L}}\!_{{\rm G}}}{{\mathscr{L}}_{{\rm I},k_{\rm t}}^{\rm t}}N\!_{\rm D} }}{2K}},
    \label{ExOmatfin}
\end{alignat}
and 
\begin{alignat}{1}
    \nonumber
    &{{\mathbb{E}}\!\left[\!\left| \left({\boldsymbol{h}_{{\rm d},k_{\rm r}}} \!\!+\! 
    {\boldsymbol{h}_{{\rm D},k_{\rm r}}^{\rm r}}\!\!(t_{D\!T})\right)^{\!H}\!{\!\boldsymbol{w}_{{\rm{\!Z\!F}},k_{\rm r}}}\!\right|^2\right]}  \ge \\
    &\;\;\;\;\;\;\;\;\;\;\;\;\; { \frac{P_0(\!N\!_{\rm A}-K)}{ {\sum\limits_{k = 1}^K \!{\frac{1}{ {{\mathscr{L}}}_{{\rm d},k} }}}  } \!+ 
    \!\frac{{P_0{{\mathscr{L}}\!_{{\rm G}}}{{\mathscr{L}}_{{\rm I},k_{\rm r}}^{\rm r}}N\!_{\rm D} }}{2K}},
    \label{ExOmarfin}
\end{alignat}

Moreover, similar to the derivations in~\eqref{Thdkt} and~\eqref{Thdkr}, 
we can directly reduce the expectations in the denominators expressed in~\eqref{JensenInet} and~\eqref{JensenIner} to
\begin{equation} 
    {\mathbb{E}}\!\!\left[{ \!\sum\limits_{u \ne {k_{\rm{t}}}}\!\! \left| \!\left(  \!{\boldsymbol{h}_{{\!\rm D},k_{\rm t}}^{\rm t}}\!\!(t_{\rm{D\!T}})\!\right)^{\!H}\!\!{\!\boldsymbol{w}_{{\rm{\!Z\!F}},u}\!}\right|^2}\!\right] 
    {\mathop  \to \limits^{\rm{p}} } 
    \frac{P_0{{{\mathscr{L}}\!_{{\rm G}}}\!\!\left(\!\sum\limits_{u \ne {k_{\rm{t}}}}\!{{\mathscr{L}}}_{{\rm I},u}\!\right)\!}{N\!_{\rm D}} }{2K},
    \label{Dent}
\end{equation}
and
\begin{equation} 
    {\mathbb{E}}\!\!\left[{ \!\sum\limits_{u \ne {k_{\rm{r}}}}\!\! \left| \!\left(  \!{\boldsymbol{h}_{{\!\rm D},k_{\rm r}}^{\rm r}}\!\!(t_{\rm{D\!T}})\!\right)^{\!H}\!\!{\!\boldsymbol{w}_{{\rm{\!Z\!F}},u}\!}\right|^2}\!\right] 
    {\mathop  \to \limits^{\rm{p}} } 
    \frac{P_0{{{\mathscr{L}}\!_{{\rm G}}}\!\!\left(\!\sum\limits_{u \ne {k_{\rm{r}}}}\!{{\mathscr{L}}}_{{\rm I},u}\!\right)\!}{N\!_{\rm D}} }{2K}.
    \label{Denr}
\end{equation}

Combining the formulations from~\eqref{ExOmatfin} to~\eqref{Denr}, we can obtain  Theorem~\ref{Theorem1}.
\end{IEEEproof}

Moreover, the more-explicit lower bounds of $R^{\rm t}_{\rm{sum}}$ and $R^{\rm r}_{\rm{sum}}$ in Theorem~\ref{Theorem2}
are derived based on the variable-amplitude IOS model.
\begin{theorem}
    \label{Theorem2}
For a variable-amplitude DIOS,
the lower bound on the ergodic achievable R\&R sum rate $R^{\rm t}_{\rm{sum}}$ and $R^{\rm r}_{\rm{sum}}$
converges in probability towards a fixed value as $N_{\rm D} \to \infty $, i.e., 
\begin{alignat}{1}
    R_{{\rm{sum}}}^{\rm t} & {\mathop  \to \limits^{\rm{p}} }  
    \!\sum\limits_{{k_{\rm t}} \in {{\cal K}_{\rm t}}}\!\!{\rm{log}_2}\!\!\left(\!\!1\!+\!\!
    { \frac{ {\mathbb{E}}\!\!\left[\!\frac{P_0}{ {\rm {tr}}{{\left( \!{{\bf{H}}_{\rm{d}}^H{{\bf{H}}_{\rm{d}}}} \right)}^{ \!- 1}}}\right] \!+ 
    \!\frac{P_0{{\mathscr{L}}\!_{{\rm G}}}{{\mathscr{L}}_{{\rm I},k_{\rm t}}^{\rm t}}N\!_{\rm D}{\mu} }{K}}
    { \frac{P_0{{{\mathscr{L}}\!_{{\rm G}}}\!\left(\!\sum\nolimits_{u \ne {k_{\rm{t}}}}\!{{\mathscr{L}}}_{{\rm I},u}\!\right)}{N\!_{\rm D}}{\mu} }{K} \!+\!\delta^2}}\!\right) \label{JensenInetLBStaChaDM}\\
    & \ge \!\sum\limits_{{k_{\rm t}} \in {{\cal K}_{\rm t}}}\!\!{\rm{log}_2}\!\!\left(\!\!1\!+\!\!
    { \frac{ \frac{P_0K(\!N\!_{\rm A}-K)}{ \!{\sum\nolimits_{k = 1}^K \!{\frac{1}{ {{\mathscr{L}}}_{{\rm d},k} }}}  } \!+ 
    \!{P_0{{\mathscr{L}}\!_{{\rm G}}}{{\mathscr{L}}_{{\rm I},k_{\rm t}}^{\rm t}}N\!_{\rm D}{\mu} }}
    {{P_0{{{\mathscr{L}}\!_{{\rm G}}}\!\left(\!\sum\nolimits_{u \ne {k_{\rm{t}}}}\!{{\mathscr{L}}}_{{\rm I},u}\!\right)}{N\!_{\rm D}}{\mu} } \!+\!K\delta^2}}\!\right),
\label{JensenInetLBDM}
\end{alignat}
and 
\begin{alignat}{1}
    R_{{\rm{sum}}}^{\rm r} & {\mathop  \to \limits^{\rm{p}} }  \!
    \!\sum\limits_{{k_{\rm r}} \in {{\cal K}_{\rm r}}}\!\!\! {\rm{log}_2}\!\!\left(\!\!1\!+\!\!
    { \frac{ {\mathbb{E}}\!\!\left[\!\frac{P_0}{{\rm {tr}}{{\left( \!{{\bf{H}}_{\rm{d}}^H{{\bf{H}}_{\rm{d}}}} \right)}^{ \!- 1}}}\!\right] \!\!+ 
    \!\frac{P_0{{\mathscr{L}}\!_{{\rm G}}}{{\mathscr{L}}_{{\rm I},k_{\rm r}}^{\rm r}}\!N\!_{\rm D}({1-\mu}){N\!_{\rm A}}}{K}}
    { \frac{P_0{{{\mathscr{L}}\!_{{\rm G}}}\!\left(\!\sum\nolimits_{u \ne {k_{\rm{r}}}}\!{{\mathscr{L}}}_{{\rm I},u}\!\right)}{N\!_{\rm D}}({1-\mu}) }{K} \!+\!\delta^2}}\!\right) \label{JensenInerLBStaChaDM}\\
    & \ge\! \!\!\sum\limits_{{k_{\rm t}} \in {{\cal K}_{\rm t}}}\!\!\!{\rm{log}_2}\!\!\!\left(\!\!1\!\!+\!\!
    { \frac{ \frac{P_0K(\!N\!_{\rm A}-K)}{ \!{\sum\nolimits_{k = 1}^K \!{\frac{1}{ {{\mathscr{L}}}_{{\rm d},k} }}}  } \!+ 
    \!{P_0{{\mathscr{L}}\!_{{\rm G}}}{{\mathscr{L}}_{{\rm I},k_{\rm r}}^{\rm r}}\!N\!_{\rm D}(1\!-\!\mu){N\!_{\rm A}}}}
    {{P_0{{{\mathscr{L}}\!_{{\rm G}}}\!\!\left(\!\sum\nolimits_{u \ne {k_{\rm{r}}}}\!{{\mathscr{L}}}_{{\rm I},u}\!\right)}\!{N\!_{\rm D}}(1\!-\!\mu) } \!\!+\!\!K\delta^2}}\!\!\right).
\label{JensenInerLBDM}
\end{alignat}
\end{theorem}
    
\begin{IEEEproof}
    The proof of Theorem~\ref{Theorem2} is similar to those of Theorem~\ref{Theorem1}.
    The main difference, however, concerns the expectations of the absolute value squareds of 
    $\left({\boldsymbol{h}_{{\rm D},k_{\rm t}}^{\rm t}}\!\!(t_{D\!T})\!\right)\!^{H}{\!\boldsymbol{w}_{{\rm{\!Z\!F}},k_{\rm t}}}$
    and $\left({\boldsymbol{h}_{{\rm D},k_{\rm r}}^{\rm r}}\!\!(t_{D\!T})\!\right)\!^H{\!\boldsymbol{w}_{{\rm{\!Z\!F}},k_{\rm r}}}$ 
    expressed in~\eqref{Thdkt} and~\eqref{Thdkr}. 
    More specifically, 
    based on the statistical characteristics derived in Proposition~\ref{Proposition2}, 
    the expectations of $\left\| \!\left({\boldsymbol{h}_{{\rm D},k_{\rm t}}^{\rm t}}\!\!(t_{D\!T})\!\right)\!^{H}{\!\boldsymbol{w}_{{\rm{\!Z\!F}},k_{\rm t}}}\! \right\|^2$ 
    and
    $\left\| \!\left({\boldsymbol{h}_{{\rm D},k_{\rm r}}^{\rm r}}\!\!(t_{D\!T})\!\right)\!^H{\!\boldsymbol{w}_{{\rm{\!Z\!F}},k_{\rm r}}}\! \right\|^2$ are given by
    \begin{alignat}{1}
        {{\mathbb{E}}\!\!\left[\!\left| \!\left({\boldsymbol{h}_{{\rm D},k_{\rm t}}^{\rm t}}\!\!(t_{D\!T})\!\right)\!^H\!{\!\boldsymbol{w}_{{\rm{\!Z\!F}},k_{\rm t}}}\!\right|^2\right]} 
        & {\mathop  \to \limits^{\rm{p}} } 
         {{{\mathscr{L}}\!_{{\rm G}}}{{\mathscr{L}}_{{\rm I},k_{\rm t}}^{\rm t}}N\!_{\rm D}\mu {{\mathbb{E}}\!\left[\!\left\| \!{\boldsymbol{w}_{{\rm{\!Z\!F}},k_{\rm t}}}\!\right\|^2\right]} } \label{Thdktxxyy} \\
        & {\mathop  \to \limits^{\rm{p}} }  \frac{{{\mathscr{L}}\!_{{\rm G}}}{{\mathscr{L}}_{{\rm I},k_{\rm t}}^{\rm t}}N\!_{\rm D} \mu P_0 }{K},
        \label{Thdktyy}
    \end{alignat}
    and
    \begin{alignat}{1}
        {{\mathbb{E}}\!\!\left[\!\left| \!\left({\boldsymbol{h}_{{\rm D},k_{\rm R}}^{\rm r}}\!\!(t_{D\!T})\!\right)\!^H\!{\!\boldsymbol{w}_{{\rm{\!Z\!F}},k_{\rm r}}}\!\right|^2\right]} &
        {\mathop  \to \limits^{\rm{p}} } 
         {{{\mathscr{L}}\!_{{\rm G}}}{{\mathscr{L}}_{{\rm I},k_{\rm r}}^{\rm r}}N\!_{\rm D}(1\!-\!\mu) {{\mathbb{E}}\!\left[\!\left\| \!{\boldsymbol{w}_{{\rm{\!Z\!F}},k_{\rm r}}}\!\right\|^2\right]}} \label{Thdkrxxyy}\\
        &{\mathop  \to \limits^{\rm{p}} }
        \frac{{{\mathscr{L}}\!_{{\rm G}}}\!{{\mathscr{L}}_{{\rm I},k_{\rm r}}^{\rm r}}\!N\!_{\rm D}(1\!-\!\mu)\!P_0}{K},
        \label{Thdkryy}
    \end{alignat}
as ${N_{\rm{D}}} \to \infty$.
\end{IEEEproof}
   
From Theorem~\ref{Theorem1}, one can see that the omnidirectional jamming impacts of the DIOS-based FPJ
refractive-side LUs and reflective-side LUs are independent on the quantization bits and the distribution of 
the DIOS coefficients. 
According to~\eqref{JensenInetLB} and~\eqref{JensenInerLB}, the jamming impact is related to the element number of the DIOS.
However, based on Theorem~\ref{Theorem2}, one can see that the jamming impact depends on statistical parameter $\mu$.
Namely, the possible amplitude values of each DIOS element and the distribution of the DIOS amplitudes 
(i.e., the quantization bits and the distribution of DIOS phase shifts).
It can be seen that from~\eqref{JensenInetLBDM} and~\eqref{JensenInerLBDM} 
the jamming impacts on the refractive-side LUs and the reflective-side LUs 
can be tuned by adjusting $\mu$, i.e.,  the quantization bits and the distribution of the DIOS coefficients. 
Therefore, we can design a appropriate distribution to balance the impacts of the DIOS-based omnidirectional fully-passive jamming attacks on
the refractive-side LUs and the reflective-side LUs.

\section{Simulation Results and Discussion}\label{ResDis}
In this section, we present numerical results to show the impact of the proposed DIOS-based FPJ.
We consider an MU-MISO system, where a legitimate AP is equipped with 128-element antenna array to communicate with 24 single-antenna LUs, i.e., $N_{\rm A} = 128$ and $K = 24$.
Furthermore,
the AP is assumed to be located at (0m, 0m, 10m) 
and the R\&R LUs are randomly distributed in the circular region $\cal S$ centered at (0m, 180m, 0m) with a radius of 20m.
In addition, a one-bit DIOS (i.e., $b=1$)  with 2,048 elements ($N_{\rm D} = 2,048$) is deployed at (2m, 2m, 8m) 
to implement fully-passive jamming to attack these LUs. 
Without loss of generality, we assume that these LUs are uniformly distributed in $\cal S$,
with half of them on the refractive-side of the DISO and the other half on the reflective-side of the DIOS.

\begin{table}    
    \footnotesize
    \centering
    \caption{A One-Bit DIOS With Variable Amplitudes}
    \label{tab1}
    \begin{tabular}{|c|c|c|c|c|c|}
    \hline
    Index   &${{\theta}^{\rm r}_{m}}$    &${{\xi}^{\rm r}_{m}}$      &${{\theta}^{\rm t}_{m}}$    &${{\xi}^{\rm t}_{m}}$    & $P_m$\\
    \hline
    $m=1$   &${\pi}/{9}$                 &0.62                       &${5\pi}/{3}$                &0.78                     &0.25\\
    \hline
    $m=2$   &${7\pi}/{6}$                &0.57                       &${2\pi}/{3}$                &0.82                     &0.75\\
    \hline
    \end{tabular}
\end{table}

For the DIOS with variable amplitudes, its R\&R coefficients are given in Table~\ref{tab1}~\cite{IOS3}.
For the DIOS with the constant amplitude, we assume that the R\&R phase shifts are the same to those in Table~\ref{tab1},
while the amplitude is $\frac{\sqrt{2}}{2}$.
Furthermore, the probabilities of taking $\theta^{\rm t}_{1}$ and $\theta^{\rm t}_{2}$ are 0.25 and 0.75.
As a result, $\mu$ in Theorem~\ref{Theorem2} is computed as 0.66. 
Moreover, we assume that the length of the DT phase is six times of that of the RPT phase, i.e., $C=6$.

According to the 3GPP propagation model~\cite{3GPP}, the propagation parameters of the wireless channels modeled 
in~Section~\ref{ChannelModel} are described as follows:
 ${\mathscr{L}}_{{\rm {G}}} \!=\!35.6\!+\!22{\log _{10}}({d_{i}})$
 and $ {\mathscr{L}}^{\rm t}_{{\rm {I}},k}, {\mathscr{L}}^{\rm t}_{{\rm {I}},k}, {\mathscr{L}}_{{\rm {d}},k}  \!=\!32.6\!+\!36.7{\log _{10}}({d_{{i}}})$,
where $d_{i} $ is the propagation distance.
Moreover,  the AWGN variance is $\delta^2\!=\!-170\!+\!10\log _{10}\left(BW\right)$ dBm, and $BW=180$ kHz.
If not otherwise specified, these above parameters default to the values.

\begin{figure*}[!t]
    \centering 
    \subfloat{
            \includegraphics[scale=0.56]{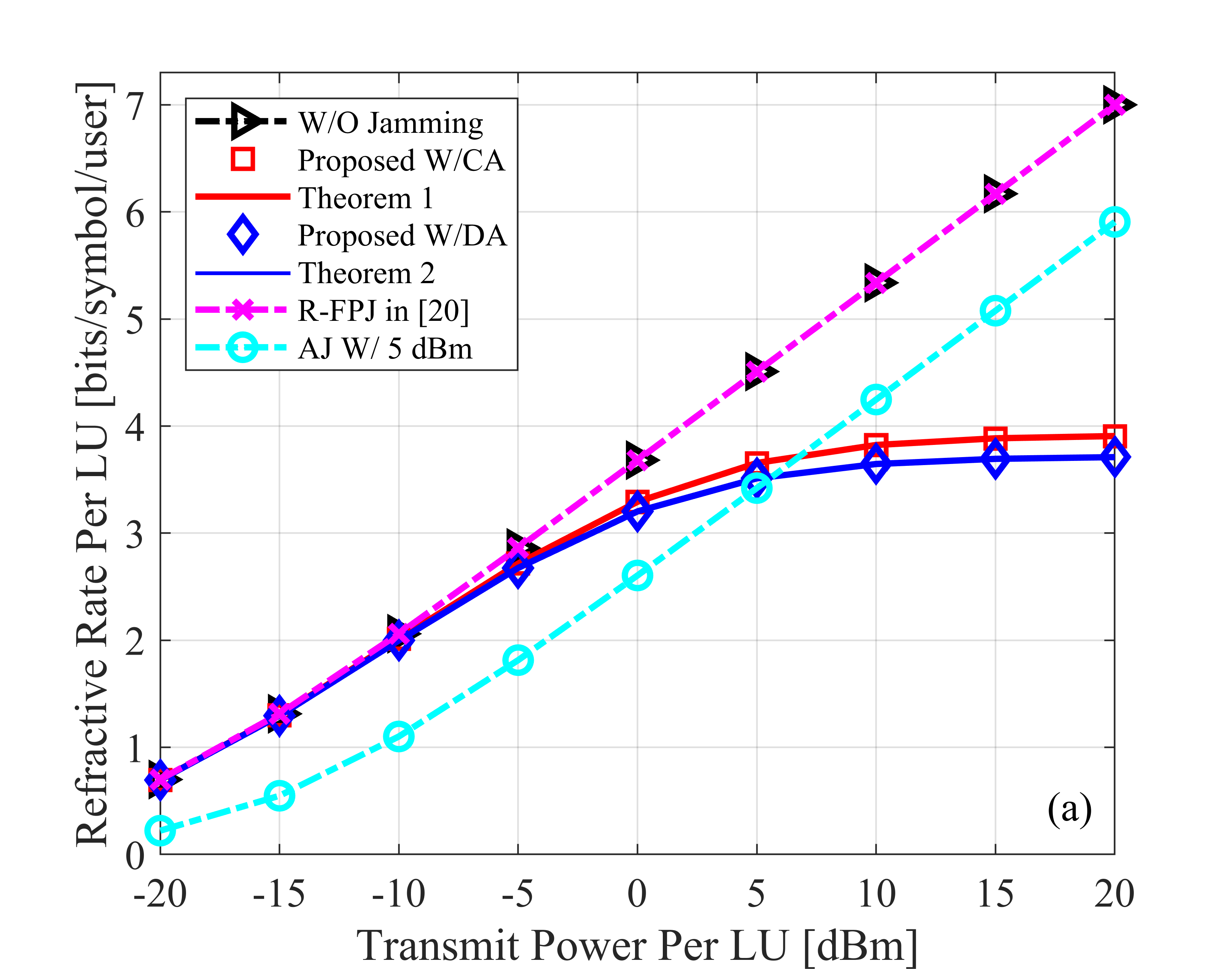}}\hspace{30pt}
    \subfloat{
            \includegraphics[scale=0.56]{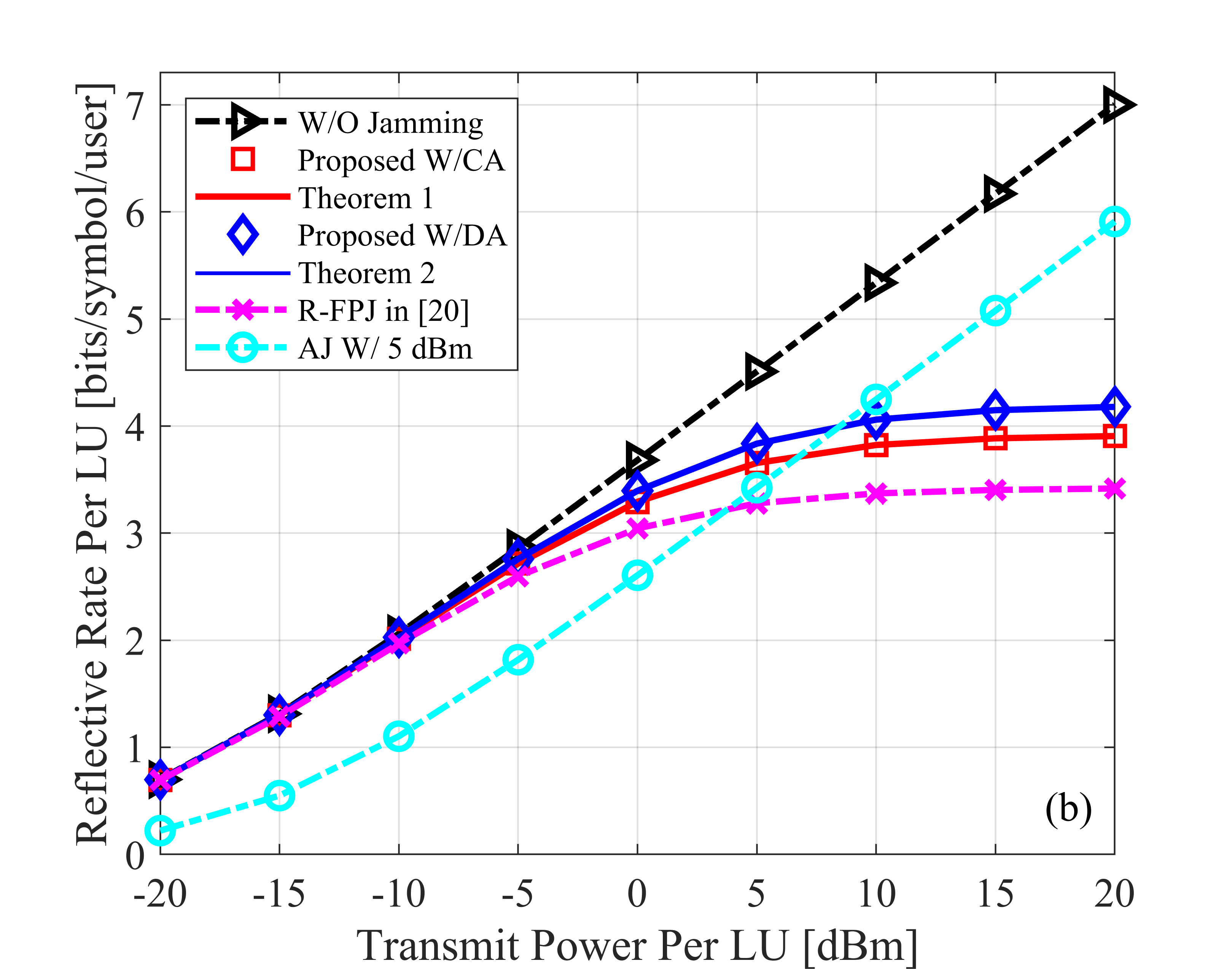}}
   \caption{Achievable performance of (a) refractive-side LUs and (b) reflective-side LUs vs transmit power for different benchmarks.}
    \label{ResFigP}
\end{figure*}

Fig.~\ref{ResFigP} shows the downlink rate per LU versus ($\frac{R_{\rm{sum}}}{K}$) the transmit power per LU ($\frac{P_0}{K}$) from different schemes 
Specifically, the performance of the following benchmarks is considered and compared:
1) the sum rate per LU without jamming attacks (W/O Jamming);
2) the sum rate per LU jammed by the proposed DIOS-based FPJ using the constant-amplitude IOS model (Proposed W/ CA) 
and 3) the corresponding theoretical analysis in Theorem~\ref{Theorem1} (Theorem~\ref{Theorem1});
4) the sum rate per LU jammed by the proposed DIOS-based FPJ using the variable-amplitude IOS model (Proposed W/ VA) 
and 5) the corresponding theoretical analysis in Theorem~\ref{Theorem2} (Theorem~2);
6) the sum rate per LU jammed by the reflective DRIS-based FPJ in~\cite{MyDRIS2} (R-FPJ in~\cite{MyDRIS2});
7) the sum rate per LU jammed by an AJ emitting 5 dBm jamming power (AJ W/ 5 dBm).
In addition, Fig.~\ref{ResFigP}~(a) illustrates the achievable performance of the refractive-side LUs via the above benchmarks,
and Fig.~\ref{ResFigP}~(b) illustrates the achievable performance of the reflective-side LUs.

\begin{figure*}[!t]
    \centering
    \subfloat{
            \includegraphics[scale=0.56]{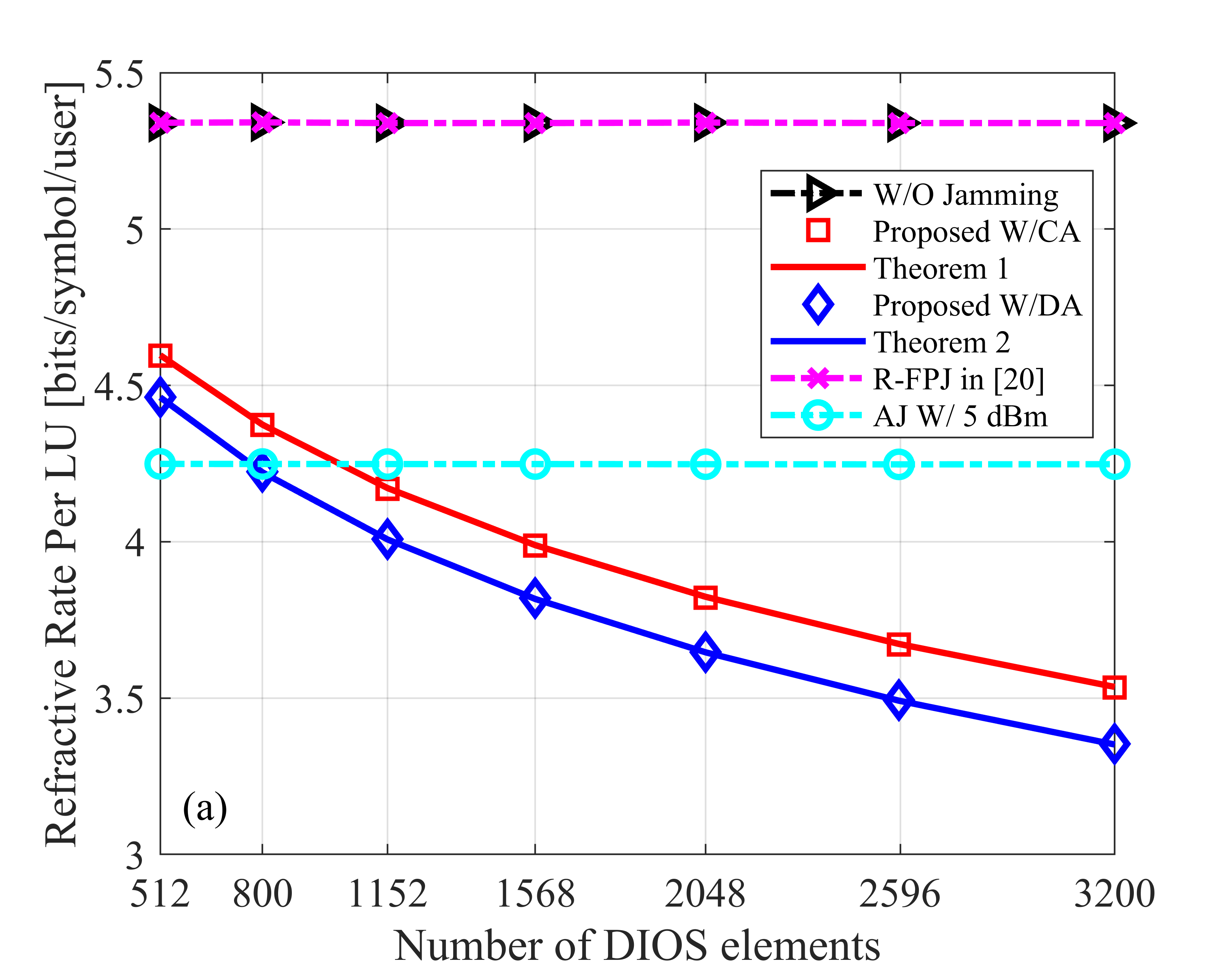}}\hspace{30pt}
    \subfloat{
            \includegraphics[scale=0.56]{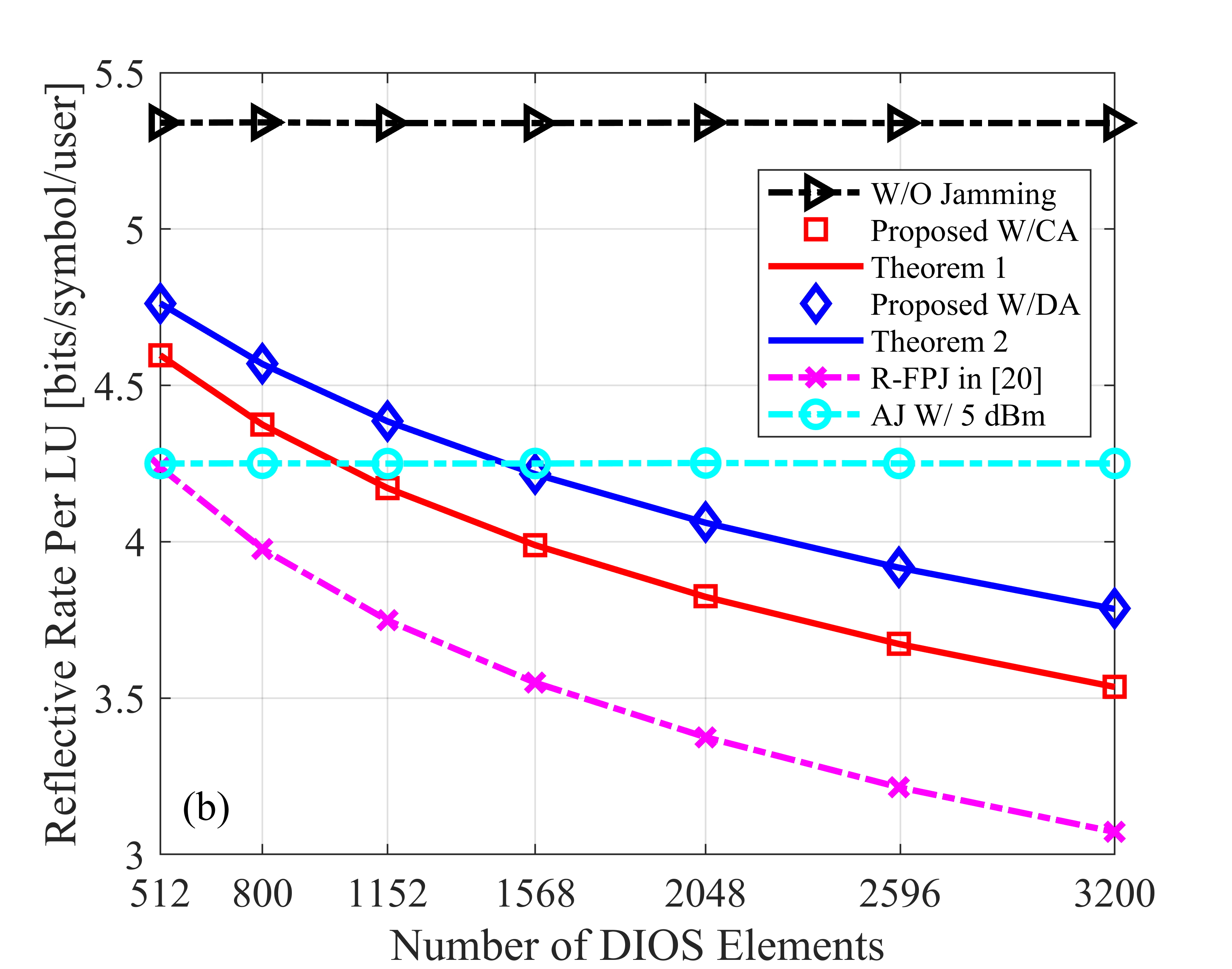}}
   \caption{Achievable performance of (a) refractive-side LUs and (b) reflective-side LUs vs the number of the DIOS elements at 10 dBm transmit power per LU for different benchmarks.}
    \label{ResFigNR}
\end{figure*}

From Fig.~\ref{ResFigP}, it can be seen that the reflective DRIS-based FPJ~\cite{MyDRIS2,MyDRIS5} jams the reflective-side LUs, but does not jam the refractive-side LUs. 
One can see that the achievable sum rate per LU of the refractive-side LUs does not decrease when attacked by the reflective DRIS-based FPJ.
However, the proposed DIOS-based FPJ can jam both the refractive-side LUs and the reflective-side LUs.
The reflective DRIS-based FPJ can achieve more severe jamming impact on the reflective-side LUs.
However, the average performance loss per LU jammed by the proposed DIOS-based FPJ is 1.5091 bits/symbol/user at 10 dBm transmit power per LU, while that caused by the DRIS-based FPJ is only 0.9746 bits/symbol/user.
In other words, the proposed DIOS-based FPJ can not only perform 360$^\circ$ fully-passive jamming,
but also improve the jamming impact by about 55\% at 10dBm transmit power per legitimate user.
Moreover, as stated aboved, $\mu$ in Theorem~\ref{Theorem2} is 0.66 based on the settings in Table~\ref{tab1}.
As a result, one can see that the jamming impact of the proposed DIOS-based FPJ on the refractive-side LUs 
is more significant than that on the reflective-side LUs. 
It is worth noting that the can change the random distribution of the DIOS phase shifts to balance the jamming effects between the refractive-side LUs and the reflective-side LUs.

Compared to the jamming impact of an AJ, the jamming impact of the proposed DIOS-based FPJ can not be suppressed by increasing
the transmit power at the legitimate AP. 
As shown in Fig.~\ref{ResFigP}, as the transmit power per LU increases, 
the jamming impact of the proposed DIOS-based FPJ increases and eventually surpasses that of the AJ.
Moreover, Fig.~\ref{ResFigP} also verifies the validity of the derived Theorem~\ref{Theorem1} and Theorem~\ref{Theorem2}.
It can be seen that the results of the asymptotic analysis provided in Theorem~\ref{Theorem1} and 
Theorem~\ref{Theorem2} are very close to the downlink rates obtained using Monte Carlo simulation.

Fig.~\ref{ResFigNR} illustrates the relationship between the achievable performance and the number of the DIOS elements.
The jamming impacts of the proposed DIOS-based FPJ on both the refractive-side LUs
and the reflective-side LUs increase with the number of the DIOS elements.
From Theorem~\ref{Theorem1} and Theorem~\ref{Theorem2},  it can also be seen that the omnidirectional
fully-passive jamming impact is caused by ACA interference, which is related to the number of the DIOS elements, i.e., $N_{\rm D}$.
Note that the proposed DIOS-based FPJ in our paper is implemented by using only 1-bit quantization DIOS phase shifts.
Therefore, it is easy to implement the proposed DIOS-based FPJ and then increase its omnidirectional jamming impact 
by using a larger number of the DIOS elements.
Although the proposed DIOS-based FPJ 

For the refractive-side LUs, the proposed DIOS-based FPJ implemented by the constant-amplitude DIOS 
can achieve the same jamming impact as the AJ with 5 dBm jamming power when the number of DIOS elements is about 1,000.
Meanwhile, when the DIOS-based FPJ is implemented by the variable-amplitude DIOS, it is enough to  achieve the same jamming impact as the AJ with 5 dBm jamming power as the number of DIOS elements is about 800.
This is because that the $\mu$ in Theorem~\ref{Theorem2} is equal to 0.66 for the refractive-side LUs,
while it is regarded as 0.5 in Theorem~\ref{Theorem1}.
As a result, the DIOS-based FPJ implemented by a variable-amplitude IOS launches more severe fully-passive jamming on the refractive-side LUs.
However, for the reflective-side LUs, the DIOS-based FPJ implemented by the variable-amplitude DIOS achieves
the same jamming impact as the AJ with 5 dBm jamming power, while the number of the DIOS elements required is 1,568.
This is because that $(1-\mu)$ is equal to 0.34 in the Theorem~\ref{Theorem2}.
As a result, the jamming impact of the DIOS-based FPJ implemented by the variable-amplitude DIOS is weaker than that of the 
DIOS-based FPJ implemented by the constant-amplitude DIOS.

\begin{figure*}[!t]
    \centering
    \subfloat{
            \includegraphics[scale=0.56]{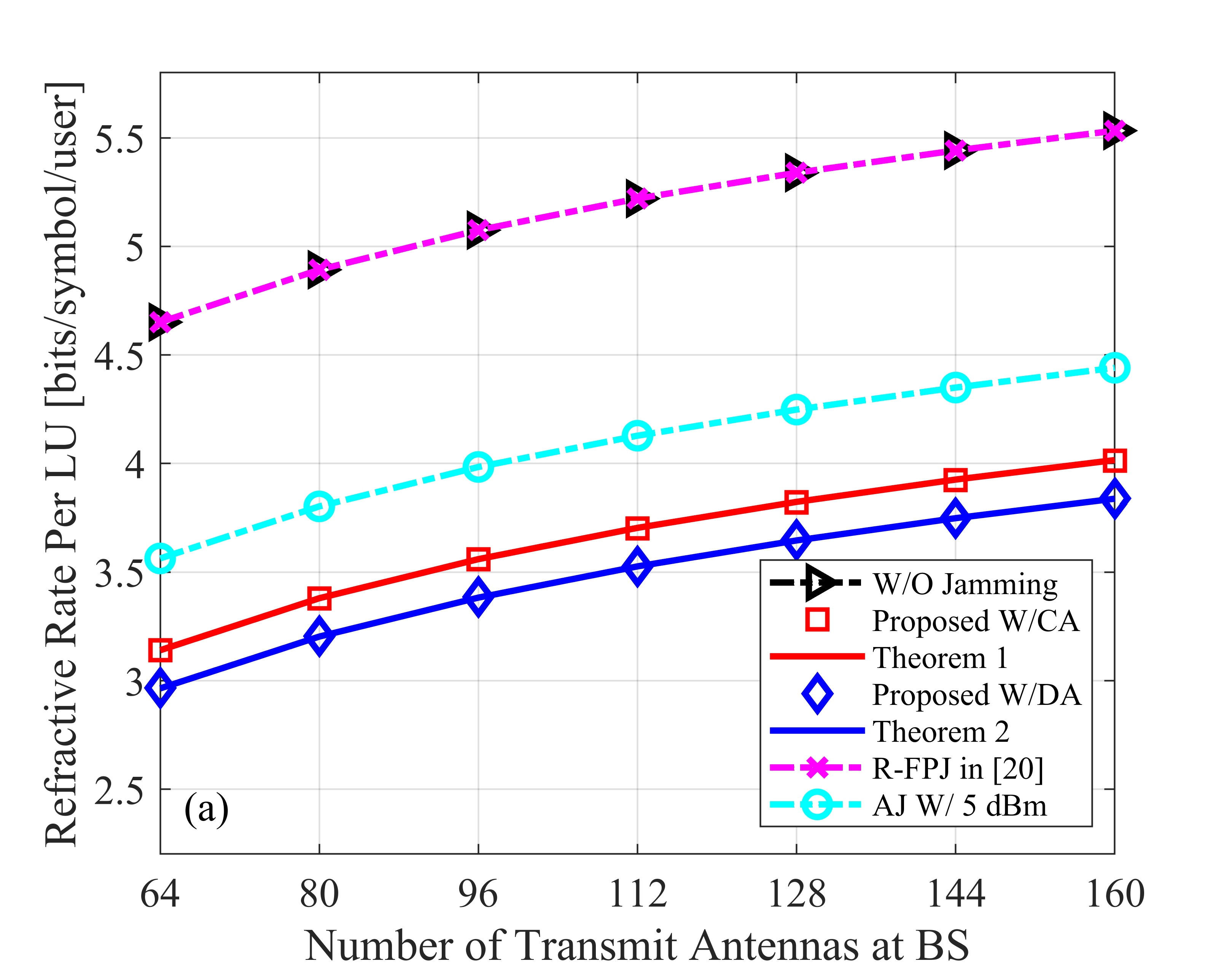}}\hspace{30pt}
    \subfloat{
            \includegraphics[scale=0.56]{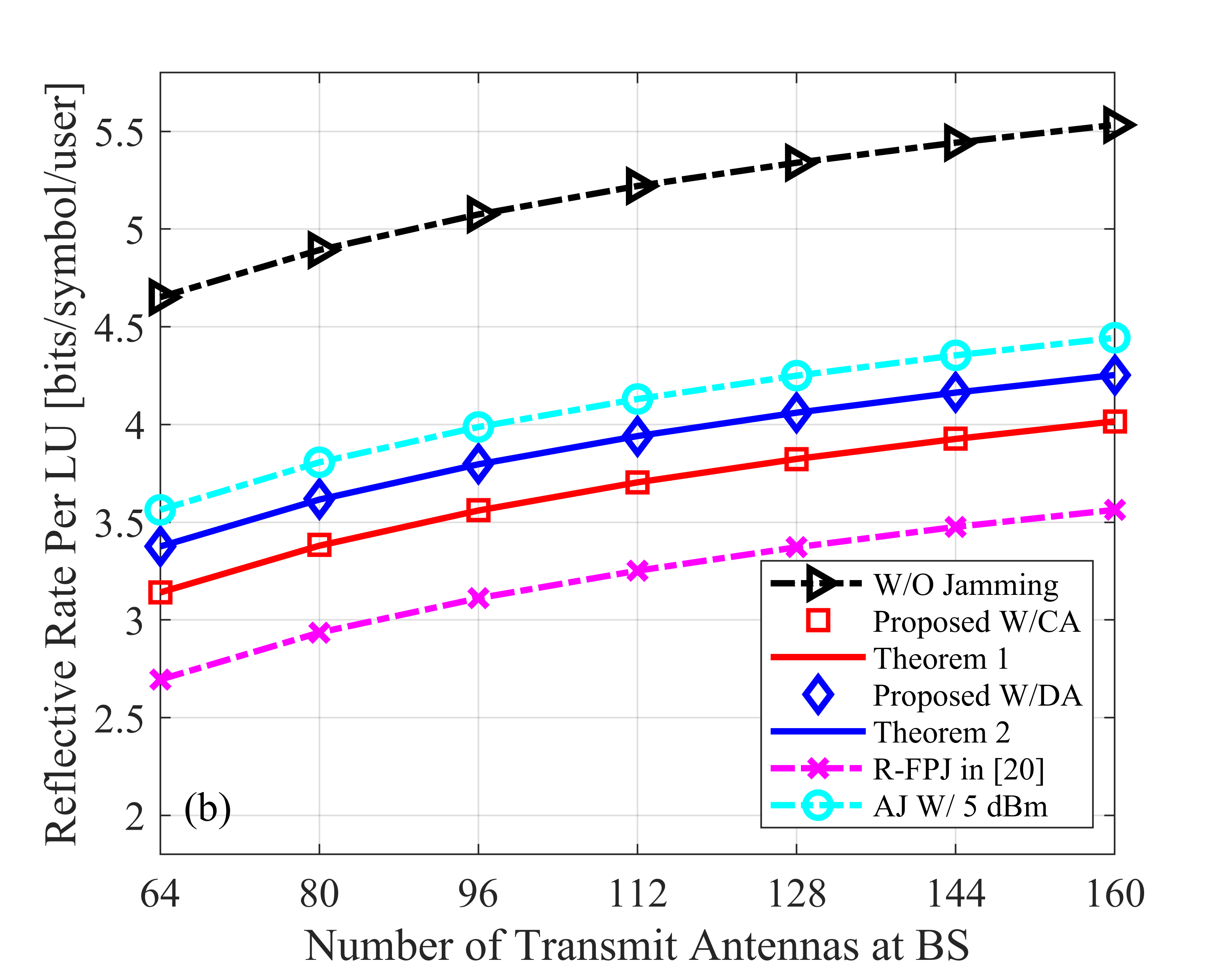}}
   \caption{Achievable performance of (a) refractive-side LUs and (b) reflective-side LUs vs the number of transmit antennas at 10 dBm transmit power per LU for different benchmarks.}
    \label{ResFigNA}
\end{figure*}

Fig.~\ref{ResFigNA} shows the performance of different benchmarks as a function of the number of transmit antennas at 10 dBm transmit power per LU.
It can be seen that the achievable sum rates per LU for both the refractive-side LUs and the reflective-side LUs
improve as the number of transmit antennas increases.
In fact, it is clear from Theorem~\ref{Theorem1} and Theorem~\ref{Theorem2} that 
the omnidirectional fully-passive jamming attacks can be suppressed by increasing $N_{\rm A}$,
which suggests that one possible scheme to mitigate the DIOS-based 360$^\circ$ fully-passive jamming is 
to use a larger number of transmit antennas at the AP.
Unfortunately, the more transmit antennas a base station has, the higher the implementation cost becomes.

\begin{figure*}[!t]
    \centering
    \subfloat{
            \includegraphics[scale=0.56]{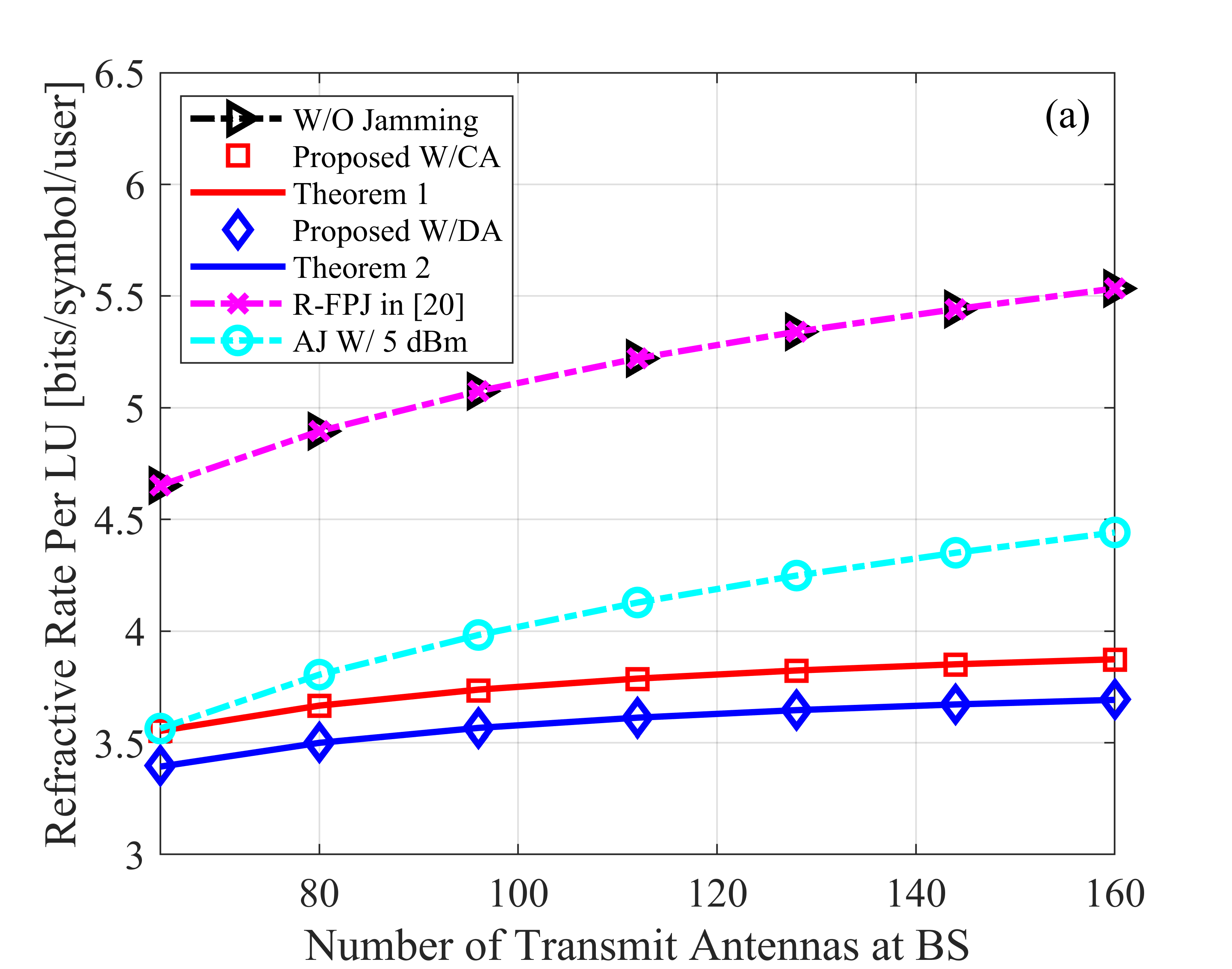}}\hspace{30pt}
    \subfloat{
            \includegraphics[scale=0.56]{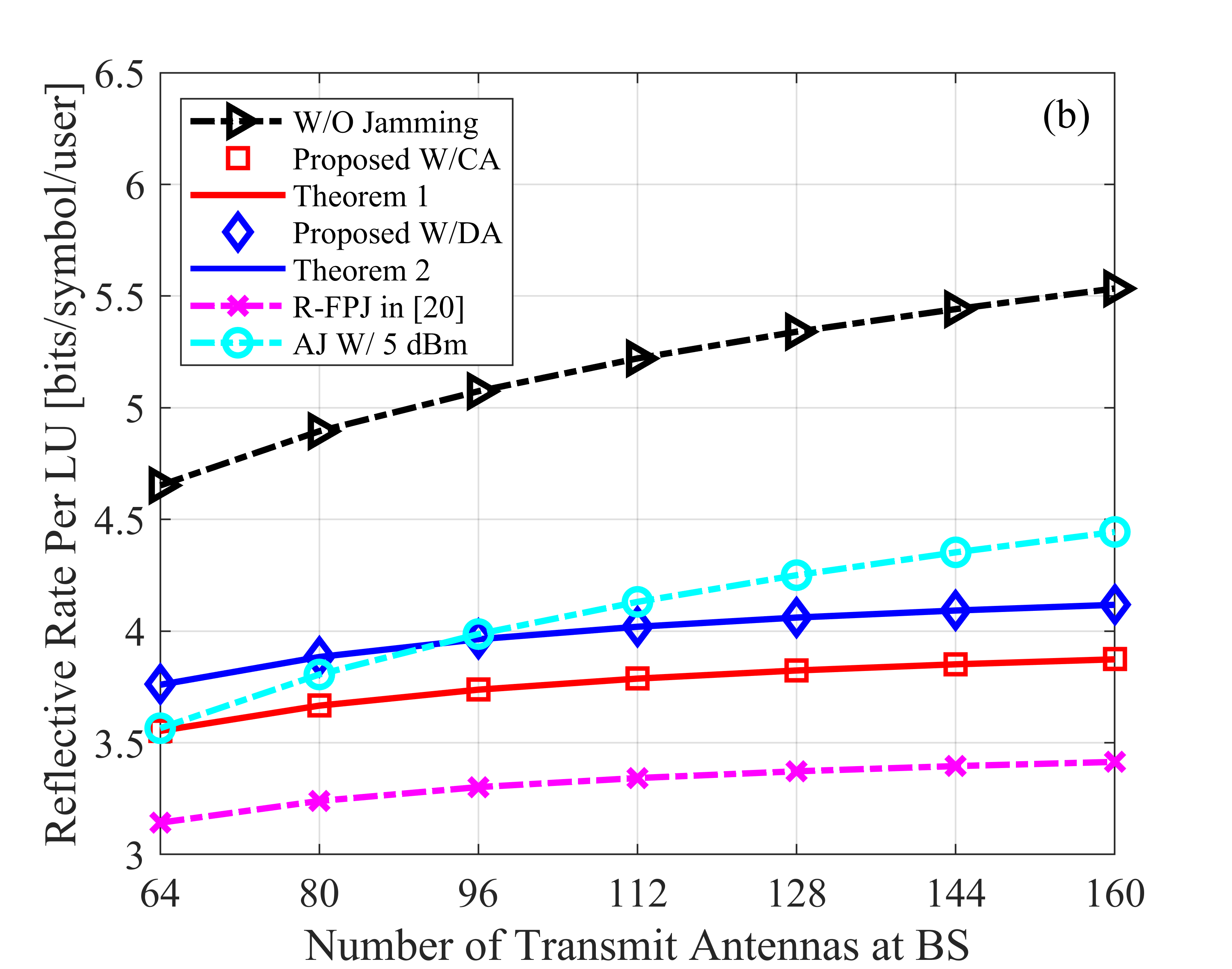}}
   \caption{Achievable performance of (a) refractive-side LUs and (b) reflective-side LUs vs the number of transmit antennas at 10 dBm transmit power per LU for different benchmarks, where the number of the DIOS is always equal to 
   the 16 times of the number of transmit antennas.}
    \label{ResFigNANR}
\end{figure*}

Moreover, the  relationship between the performance obtained from the corresponding benchmarks and the number of the AP transmit antennas,
while the number of DIOS elements also increases with the number of the transmit antennas.
Specifically, the number of the DIOS is always equal to the 16 times of the number of the transmit antennas, i.e., $N_{\rm D} = 16N_{\rm A}$.
As shown in Fig.~\ref{ResFigNANR}, an attacker can use a larger number of DIOS elements to 
counteract the mitigation provided by the increase in the AP transmit antennas.
It is worth noting that the proposed DIOS-based FPJ can be implemented by using 1-bit quantization IOS, 
which ensures that the implementation of increasing the number of DIOS elements is cheaper compared
to that of increasing the number of the transmit antennas.
Note that the traditional anti-jamming technologies, such as frequency hopping, are ineffective against the proposed DIOS-based FPJ~\cite{MyDRIS5}. 

\begin{figure*}[!t]
    \centering
    \subfloat{
            \includegraphics[scale=0.56]{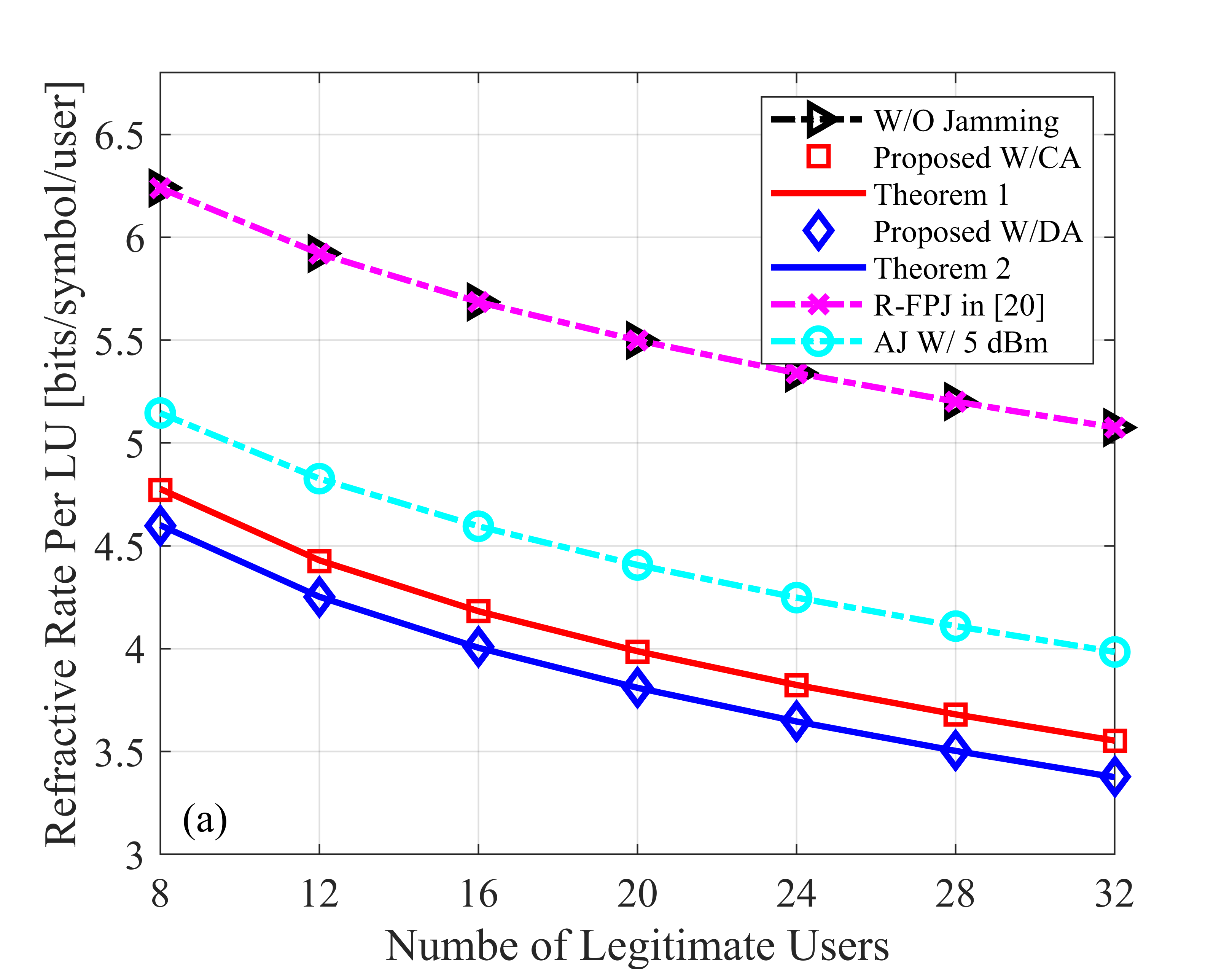}}\hspace{30pt}
    \subfloat{
            \includegraphics[scale=0.56]{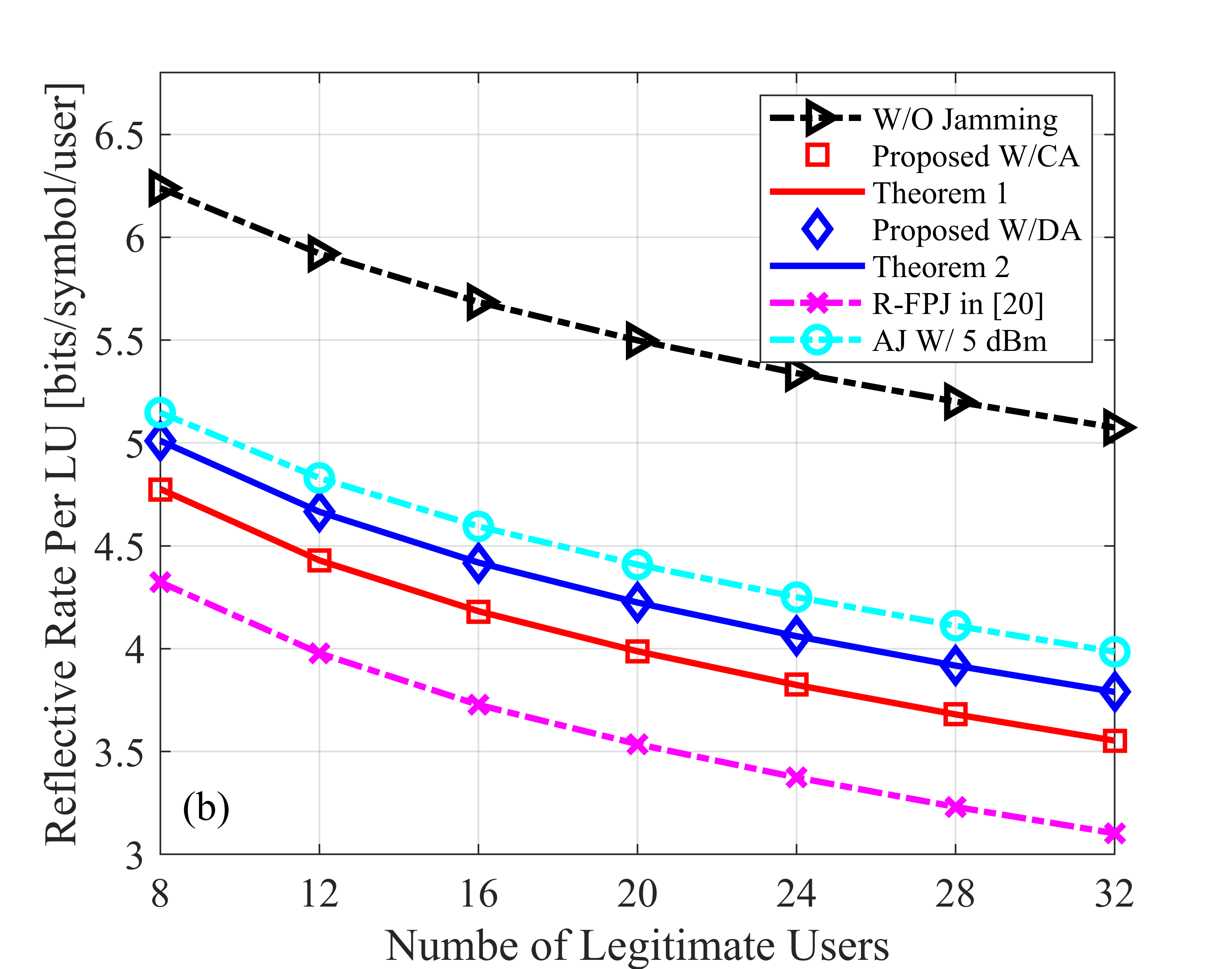}}
   \caption{Achievable performance of (a) refractive-side LUs and (b) reflective-side LUs vs the number of the LUs at 10 dBm transmit power per LU for different benchmarks.}
    \label{ResFigK}
\end{figure*}

Fig.~\ref{ResFigK} shows the relationship between the achievable performance via different benchmarks and the 
number of the LUs.
On the one hand, increasing the number of LUs reduces the gain generated by the transmit precoding at the .
On the other hand, the greater the number of LUs, the greater the ACA interference caused by the DIOS.
As a result, one can see that the omnidirectional fully-passive jamming impact becomes more severe 
as the number of the LUs increases.
In future scenarios of ultra-massive user access in 6G, 
the proposed DIOS-based FPJ poses a serious potential threat,
particularly when the number of legitimate users (LUs) is extremely large.
Therefore, it is necessary for the legitimate AP to investigate other more cost-effective anti-jamming solutions, for instance,
the anti-jamming precoding~\cite{MyDRIS3,MyDRIS4}.
\section{Conclusions}\label{Conclu}
In this work, we proposed a DIOS-based FPJ to launch 360$^{\rm o}$ fully-passive jamming attacks on MU-MISO systems.
Unlike existing AJs and RIS-based PJs, the proposed DIOS-based FPJ leverages ACA interference to launch omnidirectional fully-passive jamming attacks. 
As a result, the DIOS-based FPJ operates without requiring neither jamming power nor LU channel knowledge.
To characterize the impact of the DIOS-based FPJ on the MU-MISO system, we first derived the statistical characteristics of the DIOS-jammed channels
based on the two considered IOS model. Then, a lower bound 
of the achievable sum rates under the constant-amplitude DIOS and variable-amplitude DIOS assumptions are obtained based on the derived statistical characteristics.

The following properties are resulted from the theoretical derivations: 
1) The omnidirectional jamming impact of the proposed DIOS-based FPJ implemented by a constant-amplitude IOS does not depend on neither the quantization number nor the stochastic distribution of the DIOS coefficients;
2) However, the omnidirectional jamming impact of the proposed DIOS-based FPJ depends on the quantization bits and the stochastic distribution of the DIOS coefficients when the variable-amplitude DIOS is used. 
Therefore, we can use a variable-amplitude DIOS and  carefully design a DIOS coefficient distribution
to balance the jamming impacts on the refractive-side LUs and the reflective-side LUs.

The proposed DIOS-based FPJ can not only launch 360$^{\rm o}$ fully-passive jamming attacks,
but also achieves a more severe jamming impact compared to the existing DRIS-based FPJ.
Increasing the transmit power at the AP does not mitigate the omnidirectional jamming attacks initiated by the proposed DIOS-based FPJ; 
Instead, it exacerbates the jamming impact. 
In addition, the DIOS-based FPJ can effectively evade conventional anti-jamming techniques, including frequency hopping.
Although the APs can mitigate the proposed DIOS-based omnidirectional fully-passive jamming attacks by increasing its transmit antennas, 
this countermeasure becomes less effective as the number of DIOS elements increases.

\begin{appendices}
\section{Proof of Proposition~\ref{Proposition1}}\label{AppendixA}
Based on the definition of ${\bf H}^{\rm t}_{\rm D}\!(t)$, the element ${\left[ {\bf H}^{\rm t}_{\rm D}\!(t) \right]_{n,k_{\rm t}}}$ is written as
\begin{alignat}{1}
\nonumber
{\left[\! {\bf H}^{\rm t}_{\!\rm D}\!(t) \!\right]_{n,k_{\rm t}}} \!\!\!=& \sqrt {\!\frac{{\varepsilon}_{\! \rm G}{{\mathscr{L}}\!_{{\rm G}}}{ {\mathscr{L}}^{\rm t}_{{\rm I},k_{\rm t}}}\!N\!_{\rm D}}{{1\!+\!{\varepsilon}_{\!\rm G}} }}
 \!\!\left( \!{\frac{\sum\limits_{s = 1}^{{N_{\rm{D}}}} {\!{ {\left[ {{{\widehat { \boldsymbol{h} }}^{\rm t}_{{\rm{I}},k_{\rm t}}}} \right]}_s}\!{\alpha^{\rm t}_{s}\!(t)}{e^{j{\varphi^{\rm t}_{s}}(t)}} }}{\sqrt{\!N_{\rm D}}}} \right.\!\\
\nonumber
&\left. {\!\times {{e^{ \!- \!j\!\frac{2\pi}{\lambda}\left(\!{d_{n,s}}\!-{d_n}\right)}}}} \!\right) \!\!+\!\! \sqrt {\!\frac{{{\mathscr{L}}\!_{{\rm G}}}{{\mathscr{L}}{\rm t}_{{\rm I},k_{\rm t}}} N\!_{\rm D}}{{1\!+\!\varepsilon_{\! \rm G}} }} \!\left(\!{ {\frac{\sum\limits_{s = 1}^{{N_{\rm{D}}}} {{\!{\left[ {{{\widehat { \boldsymbol{h} }}^{\rm t}_{{\rm{I}},k_{\rm t}}}} \right]}_s}}}{\sqrt{N_{\rm D}}}} } \right.\\
&\times \left. {\!{\alpha^{\rm t}_{s}\!(t)}{e^{j{\varphi^{\rm t}_{s}}(t)}}{\left[{{\widehat {\bf{G}}}^{{\rm{NLOS}}}}\right]_{n,s} } } \right),
\label{HDele}
\end{alignat}
where ${{\left[ {{{\widehat { \boldsymbol{h} }}_{{\rm{I}},k}}} \right]}_s}$ represents the $s$-th element of ${{{\widehat { \boldsymbol{h} }}_{{\rm{I}},k}}}$ in~\eqref{HIkeq}.
Conditioned on the fact that the i.d.d. elements of ${{\bf H}^{\rm t}_{\rm{I}}(t)}$, $\boldsymbol{\varphi}_{\rm t}(t)$, and $\bf G$ are independent, we have
\begin{alignat}{1}
    \nonumber
\mathbb{E}\left[a^{\rm t}_s\right] &=\! {\mathbb{E}\!\left[{ { {{ {\left[ {{{\widehat { \boldsymbol{h} }}^{\rm t}_{{\rm{I}},k_{\rm t}}}}\!(t) \right]}_s}{ \alpha^{\rm t}_{s}(t)}{e^{j{\varphi^{\rm t}_{s}}(t)}} }} {e^{ \!- \!j\!\frac{2\pi}{\lambda}\left(\!{d_{n,s}}\!-{d_n}\right)}}  }\right]} \\
& ={\mathbb{E}\!\left[\!{ { {\left[ {{{\widehat { \boldsymbol{h} }}^{\rm t}_{{\rm{I}},k_{\rm t}}}}\!(t) \right]}_s}  }\right]}  {\mathbb{E}\!\!\left[{ { { { \alpha^{\rm t}_{s}(t)}{e^{j{\varphi^{\rm t}_{s}}(t)}} }} {e^{ \!- \!j\!\frac{2\pi}{\lambda}\left(\!{d_{n,s}}\!-{d_n}\right)}}  }\right]}\! = 0, 
\label{HDeleexpectLoS}  
\end{alignat}
and
\begin{alignat}{1}
\nonumber
\mathbb{E}\left[b^{\rm t}_s\right] & =\! {\mathbb{E}\!\left[ { {{ {\left[ {{{\widehat { \boldsymbol{h} }}^{\rm t}_{{\rm{I}},k_{\rm t}}}}\!(t) \right]}_s}{ \alpha^{\rm t}_{s}(t)}{e^{j{\varphi^{\rm t}_{s}}(t)}} }}{\left[{{\widehat {\bf{G}}}^{{\rm{NLOS}}}}\right]_{n,s} }  \right]}  \\
& ={\mathbb{E}\!\left[\!{ { {\left[ {{{\widehat { \boldsymbol{h} }}^{\rm t}_{{\rm{I}},k_{\rm t}}}}\!(t) \right]}_s}  }\right]}  {\mathbb{E}\!\!\left[{ { { \alpha^{\rm t}_{s}(t)}{e^{j{\varphi^{\rm t}_{s}}(t)}} }}{\left[{{\widehat {\bf{G}}}^{{\rm{NLOS}}}}\right]_{n,s}   }\right]} \!= 0,
\label{HDeleexpectNLoS}
\end{alignat}
where $s=1,2,\cdots,N\!_{\rm D}$. 
Since the amplitude $\alpha^{\rm t}_{s}(t), \forall s$ is assumed to be constant and equal to $\frac{\sqrt 2}{2}$, the variances of $a^{\rm t}_s$ and $b^{\rm t}_s$ 
are formulated as follows:
\begin{equation}
{\rm{Var}}\!\left[a^{\rm t}_s\right] =   \frac{1}{2}{\mathbb{E}\!\!\left[\!\left|\!{ { {\left[ {{{\widehat { \boldsymbol{h} }}^{\rm t}_{{\rm{I}},k_{\rm t}}}}\!(t) \right]}_s}  }\right|^2\right]}  =  \frac{1}{2}, 
\label{HDeleVars}
\end{equation}
and
\begin{alignat}{1}
{\rm{Var}}\!\left[b^{\rm t}_s\right]  &= \frac{1}{2}{\mathbb{E}\!\left[\!\left|\!{ { {\left[ {{{\widehat { \boldsymbol{h} }}^{\rm t}_{{\rm{I}},k_{\rm t}}}}\!(t) \right]}_s}  }\right|^2\right]}  
{\mathbb{E}\!\!\left[ \!\left|{\left[{{\widehat {\bf{G}}}^{{\rm{NLOS}}}}\right]_{n,s}   }\right|^2\right]}
\label{HDeleVars11} \\
& = \frac{1}{2}.
\label{HDeleVars12}
\end{alignat}

According to the Lindeberg-L$\acute{e}$vy central limit theorem, the random variables $\sum\nolimits_{s = 1}^{{N_{\rm{D}}}} {\frac{{{a^{\rm t}_s}}}{{\sqrt {{N_{\rm{D}}}} }}}$ and $\sum\nolimits_{s = 1}^{{N_{\rm{D}}}} {\frac{{{b^{\rm t}_s}}}{{\sqrt {{N_{\rm{D}}}} }}}$ 
in~\eqref{HDele}
converge in distribution to a normal distribution $\mathcal{CN}\!\left( {0,  \frac{1}{2}} \right)$ as $N_{\rm D} \to \infty$, i.e.,
\begin{equation}
    \sum\limits_{s = 1}^{{N_{\rm{D}}}} {\frac{{{a^{\rm t}_s}}}{{\sqrt {{N_{\rm{D}}}} }}}  \mathop  \to \limits^{\rm{d}}  \mathcal{CN}\!\left( {0,  \frac{1}{2}} \right),
    \label{HDeleVarsxx}
\end{equation}
and 
\begin{equation}
    \sum\limits_{s = 1}^{{N_{\rm{D}}}} {\frac{{{b^{\rm t}_s}}}{{\sqrt {{N_{\rm{D}}}} }}}  \mathop  \to \limits^{\rm{d}}  \mathcal{CN}\!\left( {0,  \frac{1}{2}} \right).
    \label{HDeleVarsyy}
\end{equation}

Consequently,\eqref{HDele} coverages in distribution to the following distribution:
\begin{equation}
    {\left[ {\bf H}^{\rm t}_{\!\rm D}\!(t) \right]_{n,k_{\rm t}}} \mathop \to \limits^{\rm{d}}  \mathcal{CN}\!\!\left( {0,  \frac{{{{\mathscr{L}}\!_{{\rm G}}}{{\mathscr{L}}^{\rm t}_{{\rm I},k_{\rm t}}}{N\!_{\rm D}}}}{2} } \right).
\label{HDeleDis2}
\end{equation}

Similar to the derivation from~\eqref{HDele} to~\eqref{HDeleDis2},
we have 
\begin{equation}
    {\left[ {\bf H}^{\rm r}_{\!\rm D}\!(t) \right]_{n,k_{\rm r}}} \mathop \to \limits^{\rm{d}}  \mathcal{CN}\!\!\left( {0,  \frac{{{{\mathscr{L}}\!_{{\rm G}}}{{\mathscr{L}}^{\rm r}_{{\rm I},k_{\rm r}}}{N\!_{\rm D}}}}{2} } \right).
\label{HDeleDisR}
\end{equation}

\section{Proof of Proposition~\ref{Proposition2}}\label{AppendixB}
When we consider a more practical IOS model with variable amplitudes, 
the expectations in~\eqref{HDeleexpectLoS} and~\eqref{HDeleexpectNLoS}
also hold on.
However, the variance in~\eqref{HDeleVars} is then reduced to
\begin{equation}
    {\rm{Var}}\!\left[a^{\rm t}_s\right] =  {\mathbb{E}\!\!\left[\!\left|\!{ { {\left[ {{{\widehat { \boldsymbol{h} }}^{\rm t}_{{\rm{I}},k_{\rm t}}}}\!(t) \right]}_s}  }\right|^2\right]}  
    {\mathbb{E}\!\!\left[\!\left|{  {\alpha^{\rm t}_s\!(t)}  }\right|^2\right]}.  \label{HDeleVarsDApre} 
\end{equation}
Furthermore, 
\begin{equation}
    {\mathbb{E}\!\!\left[\!\left|{  {\alpha^{\rm t}_s\!(t)}  }\right|^2\right]} \!= \!\sum\limits_{m = 1}^{{2^b}} {{P_m}{{\left( {\xi _m^{\rm{t}}} \right)}^2}} = \mu.
\label{HDeleVarTPer}
\end{equation}
Consequently, \eqref{HDeleVarsDApre} is reduced to
\begin{equation}
    {\rm{Var}}\!\left[a^{\rm t}_s\right] = \mu {\mathbb{E}\!\!\left[\!\left|\!{ { {\left[ {{{\widehat { \boldsymbol{h} }}^{\rm t}_{{\rm{I}},k_{\rm t}}}}\!(t) \right]}_s}  }\right|^2\right]}  
    = \mu.  
    \label{HDeleVarsDApre11} 
\end{equation}
Similarly, 
\begin{alignat}{1}
    \nonumber
    {\rm{Var}}\!\left[b^{\rm t}_s\right]  &= {\mathbb{E}\!\left[\!\left|\!{ { {\left[ {{{\widehat { \boldsymbol{h} }}^{\rm t}_{{\rm{I}},k_{\rm t}}}}\!(t) \right]}_s}  }\right|^2\right]}  
    {\mathbb{E}\!\!\left[\!\left|{  {\alpha^{\rm t}_s\!(t)}  }\right|^2\right]}
    {\mathbb{E}\!\!\left[ \!\left|\!{\left[{{\widehat {\bf{G}}}^{{\rm{NLOS}}}}\right]_{n,s}   }\right|^2\right]} \\
    & = \mu.
    \label{HDeleVars12VA}
\end{alignat}
    
Based on the Lindeberg-L$\acute{e}$vy central limit theorem, the element ${\left[ {\bf H}^{\rm t}_{\!\rm D}\!(t) \right]_{n,k_{\rm t}}}$
converges to the following normal distribution as $N_{\rm D}$, i.e., 
\begin{equation}
    {\left[ {\bf H}^{\rm t}_{\!\rm D}\!(t) \right]_{n,k_{\rm t}}} \mathop  \to \limits^{\rm{d}}  \mathcal{CN}\!\!\left( {0,   {{{{\mathscr{L}}\!_{{\rm G}}}{{\mathscr{L}}^{\rm t}_{{\rm I},k_{\rm t}}}{N\!_{\rm D}}\mu}} } \right).
\label{HDeleDisDMT}
\end{equation}

Moreover, for the DIOS-jammed channels of the reflective-side LUs, the element ${\left[ {\bf H}^{\rm r}_{\rm D}\!(t) \right]_{n,k_{\rm r}}}$
transfers to
\begin{alignat}{1}
    \nonumber
    {\left[\! {\bf H}^{\rm r}_{\!\rm D}\!(t) \!\right]_{n,k_{\rm r}}} \!\!\!=& \sqrt {\!\frac{{\varepsilon}_{\! \rm G}{{\mathscr{L}}\!_{{\rm G}}}{ {\mathscr{L}}^{\rm r}_{{\rm I},k_{\rm r}}}\!N\!_{\rm D}}{{1\!+\!{\varepsilon}_{\!\rm G}} }} 
    \!\!\left( \!{\frac{\sum\limits_{s = 1}^{{N_{\rm{D}}}} {\!{ {\left[ {{{\widehat { \boldsymbol{h} }}^{\rm r}_{{\rm{I}},k_{\rm r}}}} \right]}_s}\!{\alpha^{\rm r}_{s}\!(t)}{e^{j{\varphi^{\rm r}_{s}}(t)}} }}{\sqrt{\!N_{\rm D}}}} \right.\!\\
    \nonumber
    &\left. {\!\times {{e^{ \!- \!j\!\frac{2\pi}{\lambda}\left(\!{d_{n,s}}\!-{d_n}\right)}}}} \!\right) \!\!+\!\! \sqrt {\!\frac{{{\mathscr{L}}\!_{{\rm G}}}{{\mathscr{L}}{\rm r}_{{\rm I},k_{\rm r}}} N\!_{\rm D}}{{1\!+\!\varepsilon_{\! \rm G}} }} 
    \!\left(\!{ {\frac{\sum\limits_{s = 1}^{{N_{\rm{D}}}} {{\!{\left[ {{{\widehat { \boldsymbol{h} }}^{\rm r}_{{\rm{I}},k_{\rm r}}}} \right]}_s}}}{\sqrt{N_{\rm D}}}} } \right.\\
    &\times \left. {\!{\alpha^{\rm r}_{s}\!(t)}{e^{j{\varphi^{\rm r}_{s}}(t)}}{\left[{{\widehat {\bf{G}}}^{{\rm{NLOS}}}}\right]_{n,s} } } \right),
    \label{HDeleR}
\end{alignat}
Consequently,
the following random variables ${a^{\rm r}_{s}}$ and ${b^{\rm r}_{s}}$ can be defined, which are
\begin{alignat}{1}
    &a^{\rm r}_s\! =\! { { {{ {\left[ {{{\widehat { \boldsymbol{h} }}^{\rm r}_{{\rm{I}},k_{\rm r}}}}\!(t) \right]}_s}{ \alpha^{\rm r}_{s}(t)}{e^{j{\varphi^{\rm r}_{s}}(t)}} }} {e^{ \!- \!j\!\frac{2\pi}{\lambda}\left(\!{d_{n,s}}\!-{d_n}\right)}}  }  , \label{HDeleexpectLoSR} \\
    &b^{\rm r}_s\! =\! { {{ {\left[ {{{\widehat { \boldsymbol{h} }}^{\rm r}_{{\rm{I}},k_{\rm r}}}}\!(t) \right]}_s}{ \alpha^{\rm r}_{s}(t)}{e^{j{\varphi^{\rm r}_{s}}(t)}} }}{\left[{{\widehat {\bf{G}}}^{{\rm{NLOS}}}}\right]_{n,s} }.
    \label{HDeleexpectNLoSR}
\end{alignat}
Similar to the derivations of~\eqref{HDeleVarsDApre11} and~\eqref{HDeleVars12VA}, the variances of ${a^{\rm r}_{s}}$ and ${b^{\rm r}_{s}}$ is written as
\begin{equation}  
        {\rm{Var}}\left[a^{\rm r}_s\right] \!= \!{\rm{Var}}\left[b^{\rm r}_s\right]
        \!=\! {\mathbb{E}}\!\left[{\alpha^{\rm r}_s\!(t)}\!\left(\alpha^{\rm r}_s\!(t)\right)^{\! H}\right] \!= \!\!\sum\limits_{m = 1}^{{2^b}} {{P_m}{{\left( {\xi _m^{\rm{r}}} \right)}^2}}.
    \label{HDeleVarsR}
\end{equation}
Note that, due to the energy constraint of an IOS, we have ${{\left( {\xi _m^{\rm{t}}} \right)}^2} = 1-{{\left( {\xi _m^{\rm{r}}} \right)}^2}, \forall m$.
Therefore,~\eqref{HDeleVarsR} reduces to 
\begin{equation}
    {\rm{Var}}\left[a^{\rm r}_s\right] = {\rm{Var}}\left[b^{\rm r}_s\right] = 1-\mu.
\label{HDeleDisDMR}
\end{equation}
Based on the Lindeberg-L$\acute{e}$vy central limit theorem, the element ${\left[ {\bf H}^{\rm r}_{\!\rm D}\!(t) \right]_{n,k_{\rm r}}}$
follows a normal distribution. Specifically,
\begin{equation}
    {\left[ {\bf H}^{\rm r}_{\!\rm D}\!(t) \right]_{n,k_{\rm r}}} \mathop  \to \limits^{\rm{d}}  \mathcal{CN}\!\!\left( {0,   {{{{\mathscr{L}}\!_{{\rm G}}}{{\mathscr{L}}^{\rm r}_{{\rm I},k_{\rm r}}}{N\!_{\rm D}}({1-\mu})}} } \right).
\label{HDeleDisDMR1}
\end{equation}
\end{appendices}

\end{document}